\documentclass[aps,showpacs,amsfonts,nofootinbib,preprintnumbers,nobalancelastpage]{revtex4}

\usepackage[normalem]{ulem}
\usepackage[T1]{fontenc}
\usepackage{slashed,amsmath,amsfonts,amssymb,bbm}
\usepackage{mathrsfs,wasysym}
\usepackage{units}
\usepackage{hyperref}
\usepackage{graphicx}
\usepackage{braket}
\usepackage[dvipsnames]{xcolor}

\newcommand{\tpdf}[2]{\texorpdfstring{#1}{#2}}

\newcommand{\g}{\gamma}
\renewcommand{\a}{\alpha}
\newcommand{\de}{\partial}
\newcommand{\hc}{\rm h.c.}
\renewcommand{\O}{\mathcal{O}}

\newcommand{\CB}{C_{\tilde B}}
\newcommand{\CW}{C_{\tilde W}}
\newcommand{\CG}{C_{\tilde G}}
\newcommand{\CaPhitwo}{C_{a\Phi}^{(2)}}

\begin{document}

\title{Unitarity constraints on ALP interactions}
\author{I.~Brivio,}
\email{brivio@thphys.uni-heidelberg.de}
\affiliation{Institut f\"ur Theoretische Physik, Universit\"at
  Heidelberg.\\
  Philosophenweg 16, 69120 Heidelberg (Germany)}

\author{O.~J.~P. \'Eboli,}
\email{eboli@if.usp.br}
\affiliation{Instituto de F\'isica, Universidade de S\~{a}o Paulo,
  S\~{a}o Paulo -- SP 05508-090, Brazil}

\author{M.~C.~Gonz\'alez-Garc\'ia}
\affiliation{Instituci\'o Catalana de Recerca i Estudis Avan\c cats
  (ICREA),\\  Departament d'Estructura i Constituents de la Mat\`eria,
  Universitat de Barcelona,\\
  647 Diagonal, E-08028 Barcelona, Spain}
\affiliation{C.N. Yang Institute for Theoretical Physics, SUNY at
  Stony Brook,\\
  Stony Brook, NY 11794-3840, USA}
\email{maria.gonzalez-garcia@stonybrook.edu}

\begin{abstract}
  We derive partial-wave unitarity constraints on gauge-invariant
  interactions of an Axion-Like Particle (ALP) up to dimension-6 from
  all allowed $2\to2$ scattering processes in the limit of large
  center-of-mass energy.  We find that the strongest bounds stem from
  scattering amplitudes with one external ALP and only apply to the
  coupling to a pair of $SU(2)_L$ gauge bosons. Couplings to $U(1)_Y$
  and $SU(3)_C$ gauge bosons and to fermions are more loosely
  constrained.
\end{abstract}

\preprint{YITP-SB-2021-8}
\maketitle
\renewcommand{\baselinestretch}{1.15}
\section{Introduction}

Axion-like Particles (ALPs) are generic pseudo-Goldstone bosons, that
can emerge from the spontaneous breaking of some global symmetry at
energies well above the electroweak (EW) scale $v$. While the main
representative is the QCD axion (either
"invisible"~\cite{Peccei:1977hh, Peccei:1977ur, Weinberg:1977ma,
  Wilczek:1977pj, Kim:1979if, Shifman:1979if, Dine:1981rt,Zhitnitsky:1980tq} 
or in modern setups where the tie between axion
mass and couplings is relaxed~\cite{Rubakov:1997vp, Berezhiani:2000gh,
  Hsu:2004mf, Hook:2014cda, Fukuda:2015ana, Chiang:2016eav,
  Gherghetta:2016fhp, Dimopoulos:2016lvn, Kobakhidze:2016rwh,
  Agrawal:2017ksf, Agrawal:2017evu, Gaillard:2018xgk,
  Buen-Abad:2019uoc, Csaki:2019vte, Hook:2019qoh, Gherghetta:2020ofz,
  Hook:2018jle, DiLuzio:2021pxd, DiLuzio:2021gos}), this class of
particles encompasses a large number of exotic states, that can emerge
in composite Higgs models~\cite{Gripaios:2009pe,Gripaios:2016mmi,
  Merlo:2017sun, Chala:2017sjk, Brivio:2017sdm}, models with
spontaneous breaking of lepton number
("Majorons")~\cite{Chikashige:1980ui, Gelmini:1980re}, dynamical
flavor theories ("axiflavons")~\cite{,Davidson:1981zd,Wilczek:1982rv, Calibbi:2016hwq,Ema:2016ops},
string theory~\cite{Svrcek:2006yi, Arvanitaki:2009fg}
and many other scenarios. \smallskip

ALPs are usually studied within a model-independent Effective Field
Theory (EFT) framework~\cite{Georgi:1986df,Choi:1986zw}.  Their
pseudo-Goldstone nature justifies the assumption that ALPs are the
only light remnant of a much heavier new physics sector, whose
interactions are suppressed by a characteristic scale $f_a\gg v$.  A priori, the
allowed parameter space spans several orders of magnitude both in the
ALP mass $m_a$ and in the couplings to Standard Model (SM) particles
that, within the EFT approach, enter at lowest order as dimension 5
operators. \smallskip

The interest in the ALP Lagrangian as a self-consistent EFT has grown
recently, leading to several studies of its theoretical properties.
For instance, the Renormalization Group (RG) evolution and the
matching to the ALP EFT valid below the EW scale were derived
in~\cite{Choi:2017gpf, Chala:2020wvs, Bauer:2020jbp}. The interplay
between dimension-5 ALP interactions and dimension-6 operators in the
Standard Model EFT (SMEFT) was explored in~\cite{Galda:2021hbr}.  The
matching of the ALP EFT to concrete QCD axion models was examined in
Ref.~\cite{Alonso-Alvarez:2021ett}, that pointed out theoretical subtleties 
when applying the EFT approach to loop processes.
\smallskip

In this work we examine the validity range of the ALP EFT
at high energies on the basis of its perturbative partial-wave
unitarity properties.  It is well known that classically
non-renormalizable interactions give rise to rapid growth of the
scattering amplitudes with energy, which leads to partial-wave
unitarity violation at some large value of the center-of-mass energy
$\sqrt{S}$. Generically, partial-wave unitarity violation signals
the breakdown of the low-energy description and indicates that extra
fundamental degrees of freedom or the onset of the nonperturbative
regime must be present around or below the apparent unitarity
violation scale in order to restore the physical behavior of
scattering amplitudes.  Paradigmatic examples of the use of
unitarity relations to derive constraints on the validity of a
theory include the seminal work of Lee, Quigg and Thacker
\cite{Lee:1977yc,Lee:1977eg} that imposed an upper bound on the
Higgs mass by analyzing the unitarity of the standard model and was
used to build a case in favor of the construction of the present
generation of colliders. Another classical example are the bounds on
new fermions obtained by Chanowitz, Furman and Hinchliffe
\cite{Chanowitz:1978uj}.  On a more formal front, unitarity arguments
have also been employed, for example, in connection with the
requirement of gauge invariance~\cite{Cornwall:1974km}. \smallskip

In the last decades, partial-wave unitarity has been employed
ubiquitously to constrain effective interactions, in particular in
the electroweak sector (see for example ~\cite{Bilchak:1987cp,
Gounaris:1993fh, Gounaris:1994cm, Gounaris:1995ed,
Degrande:2013mh,Baur:1987mt,Dahiya:2013uba,Ghosh:2017coz}).
Recently, Refs.~\cite{Corbett:2014ora, Corbett:2017qgl,
Almeida:2020ylr} presented a general systematic study of unitarity
bounds for the case of effective interactions in the SMEFT and Higgs
EFT (HEFT). Generically, unitarity preservation imposes consistency
conditions on the theory such that, for the EFT to be valid up to a
given $\sqrt{S}$, the effective couplings (scale) need to be smaller
(larger) than a certain threshold.  Conversely, for given values of
the EFT coefficients and scale, unitarity imposes an upper limit on
the energy scales at which the EFT can be applied. In that respect
unitarity bounds are crucial for the interpretation of actual
experiments, which study tails of kinematical distributions, since
one can infer unphysical bounds that are too strong if these limits
are not respected. \smallskip

For the case of ALP EFT, the rapid growth of the scattering
amplitudes with energy, that leads to partial-wave unitarity
violation, is particularly enhanced. The reason for that is the pseudo-Goldstone
nature of the ALPs which requires all their interactions to be classically
invariant under shifts $a(x)\mapsto a(x) + \a$, \emph{i.e.} to be of the
form $J^\mu \de_\mu a $. As a consequence, an explicit momentum
dependence is present in all ALP couplings. \smallskip

A partial analysis of unitarity constraints on ALP couplings was
presented in Refs.~\cite{Marciano:2016yhf,Cornella:2019uxs}. Here we
adopt a more systematic approach and derive maximal constraints on all
ALP interactions of dimension 5 and 6 from partial-wave unitarity,
examining all allowed $2\to2$ scattering processes in the limit of
large center-of-mass energy. We adopt a procedure analogous to the one
employed in~\cite{Corbett:2014ora, Corbett:2017qgl, Almeida:2020ylr}
for the case of effective interactions in the SMEFT and HEFT.
\smallskip

The outline of this article is as follows. We present the relevant
Lagrangian employed in Sec.~\ref{sec:forma} and briefly discuss the
number of relevant operators we consider. The core of the results is
contained in Sec.~\ref{sec:results} where we derive first the most
general bounds for the ALP couplings to SM gauge bosons which  are
obtained from the partial-wave analysis of the scattering of boson
pairs in Sec.~\ref{sec:bosons}.  Section~\ref{sec:fermions} contains
our derivation of the most general independent constraints on ALP
couplings to SM fermions which are obtained with the partial-wave
analysis of scatterings involving fermion pairs.  We briefly discuss
the results in Sec.~\ref{sec:conclu}.  Explicit expressions of the
helicity amplitudes for all the relevant processes are presented in
the Appendix. \smallskip

\section{ALP effective Lagrangian}
\label{sec:forma}
We consider the SM extended by the ALP effective
Lagrangian~\cite{Georgi:1986df, Choi:1986zw, Brivio:2017ije}
\begin{equation}\label{eq:L_alp}
\begin{aligned}
\mathcal{L}_{ALP} &= \frac{1}{2} \de_\mu a\de^\mu a -
\frac{m_a^2}{2}a^2 + \CB\,O_{\tilde B} + \CW\, O_{\tilde W} + \CG\,
O_{\tilde G} + C_{a\Phi} \, O_{a\Phi} \\ & + \big[
  C_{u\Phi}\,O_{u\Phi} + C_{d\Phi}\,O_{d\Phi} + C_{e\Phi}\,O_{e\Phi}
  +\hc\big]\\
  &+ \CaPhitwo O_{a\Phi}^{(2)}\,,
\end{aligned}
\end{equation}
where the effective operators
\begin{align}
\label{eq:operators_1}
O_{\tilde B} &=\frac{a}{f_a}B_{\mu\nu}\tilde B^{\mu\nu} \,, 
&
O_{u\Phi}&=i\,\frac{a}{f_a}\;\bar q\, Y_u \tilde\Phi\, u\,, 
\\ 
O_{\tilde W} &=\frac{a}{f_a}W^i_{\mu\nu}\tilde W^{i\mu\nu} \,, 
& 
O_{d\Phi }&= i\,\frac{a}{f_a}\;\bar q\, Y_d \Phi\, d\,, 
\\ 
O_{\tilde G} &=\frac{a}{f_a}G^a_{\mu\nu}\tilde G^{a\mu\nu} \,, 
& 
O_{e\Phi}&=i\,\frac{a}{f_a}\;\bar l\, Y_e \Phi\, e\,, 
\\ 
O_{a\Phi} &= \,\frac{\de_\mu a}{f_a} (i\Phi^\dag (D_\mu\Phi)
- i (D_\mu\Phi)^\dag\Phi)
&
O_{a\Phi}^{(2)} &= \frac{\de_\mu a\de^\mu a}{f_a^2}(\Phi^\dag \Phi)
\,,
\label{eq:operators_2}
\end{align}
form a complete basis of CP-even ALP interactions up to $\O(f_a^{-3})$ terms.
Here, $B_\mu, W_\mu^i$ and $G_\mu^a$ are the gauge bosons of the
$U(1)_Y\times SU(2)_L\times SU(3)_c$ SM symmetry respectively, and the
dual field strengths are defined by
$\tilde X_{\mu\nu} = \frac{1}{2}\varepsilon_{\mu\nu\rho\sigma}
X^{\rho\sigma}$.  $\Phi$ denotes the $SU(2)_L$ Higgs doublet while
$\tilde \Phi = i\tau^2\Phi^*$ is its dual (being $\tau^i$ the Pauli
matrices). Upon EW symmetry breaking, $\langle \Phi^\dag\Phi\rangle = (v+H)^2/2$ 
with $H$ the physical Higgs boson. The left (right)-handed fermion multiplets
are denoted by $q, l$ ($u, d, e$) and $Y_u,Y_d,Y_e$ are the $3\times3$
Yukawa matrices. All index contractions were left implicit and
repeated indices are summed over unless otherwise specified.  A mass
term $m_a$ for the ALP was introduced, which is generically induced in
the presence of soft breaking of shift-invariance, such as
non-perturbative instanton effects in the case of the QCD
axion~\cite{Bardeen:1978nq,Shifman:1979if,DiVecchia:1980yfw,diCortona:2015ldu}. \smallskip

We neglect CP violating effects such that all Wilson coefficients
$C_i$ are real scalar quantities.  Although this is not manifest in
Eqs.~\eqref{eq:operators_1}--\eqref{eq:operators_2}, all ALP
interactions are classically shift-invariant:
the interactions to
bosons can be written as $\de_\mu a J^\mu_X$ by integration by parts,
where $J_X^\mu$ is the Chern-Simons current associated to the
$X=\{B,W,G\}$ gauge boson.\footnote{In the $G$ case, only a
  discrete version of the shift-invariance is preserved due to the
  presence of non-vanishing instanton configurations.}  The operators
with fermions were taken to follow the minimal-flavor-violation
ansatz~\cite{Chivukula:1987py,Hall:1990ac,D'Ambrosio:2002ex}, {\em
  i.e.} to respect a $U(3)^5$ global symmetry that is only broken by
insertions of the Yukawa couplings. With this flavor structure, they
could also be equivalently traded for a set of chirality-conserving
ones of the form $(\de_\mu a)(\bar \psi_p \g^\mu \psi_r)\delta^{pr}$,
with $p,r$ flavor indices~\cite{Chala:2020wvs,Bonilla:2021ufe}.  \smallskip

The operator $O_{a\Phi}$ is actually redundant~\cite{Georgi:1986df,
  Brivio:2017ije, Bauer:2020jbp,Bonilla:2021ufe}:
\begin{equation}
O_{a\Phi} = O_{u\Phi} - O_{d\Phi} - O_{e\Phi} +\hc \,.
\end{equation}
Nevertheless, it is often retained because the set
$\{O_{\tilde B}, O_{\tilde W}, O_{\tilde G}, O_{a\Phi}\}$ forms a
complete and non-redundant operator basis at dimension 5 in the bosonic sector that
can be of phenomenological interest. \smallskip

The operator $O_{a\Phi}^{(2)}$ has been previously considered
in~\cite{Draper:2012xt,Bauer:2017ris,Bauer:2018uxu,Davoudiasl:2021haa}
and it is the only shift-invariant operator\footnote{One more operator
  structure is present at dimension 6, namely $(\de_\mu \de^\mu
  a)^2$. However, applying the ALP equation of motion, this can be
  fully reabsorbed into a redefinition of the ALP mass. We have
  checked the completeness of the dimension-6 set with {\tt
    BasisGen}~\cite{Criado:2019ugp}.} that can be constructed at
dimension 6.  SMEFT operators of dimension 6 are neglected: we
assume them to be suppressed by a scale $\Lambda_{\rm SMEFT}\neq f_a$ and
work consistently at order $(f_a^{-2}\,\Lambda_{\rm SMEFT}^0)$.
Discussing the interplay of the two expansions is beyond the scope of
this work.\footnote{As will become
  clear from the discussion in Sec.~\ref{sec:results}, the bounds on
  $\CW$, $C_{f\Phi}$, $C_{a\Phi}$ and $\CaPhitwo$ are not expected to
  change significantly in the presence of dimension 6 SMEFT operators,
  independently of the interplay between the SMEFT and ALP
  expansions. This is because all these bounds are dominated by
  scatterings with one or two external ALPs.  }

\section{Analysis of unitarity constraints}
\label{sec:results}

\subsection{Helicity amplitudes for the scattering of pairs of bosons}
\label{sec:bosons}

Consider the two-to-two scattering of bosons $V_i$ with helicities $\lambda_i$
\begin{equation}
{V_1}_{\lambda_1}{V_2 }_{\lambda_2} \to {V_3}_{\lambda_3}{V_4}_{\lambda_4}\,,
\end{equation}
where we denote by $V$ either gauge bosons, Higgs or ALP.  The
corresponding helicity amplitude can be expanded in partial waves in
the center--of--mass system as~\cite{Jacob:1959at}
\begin{align}
\label{eq:helamp}
\mathcal{M}
({V_1}_{\lambda_1}{V_2 }_{\lambda_2} \to {V_3}_{\lambda_3}{V_4}_{\lambda_4}) 
=16 \pi \sum_J
\left (2 J+1 \right)&~ 
\sqrt{1+\delta_{{V_1}_{\lambda_1}}^{{V_2}_{\lambda_2}}}
\sqrt{1+\delta_{{V_3}_{\lambda_3}}^{{V_4}_{\lambda_4}}}
\,\times
\\
&\times
d_{\lambda\mu}^{J}(\theta) ~e^{i M \varphi}
~ T^J({V_1}_{\lambda_1}{V_2 }_{\lambda_2} \to {V_3}_{\lambda_3}{V_4}_{\lambda_4}) 
\;\;,
\nonumber
\end{align}
where $\lambda=\lambda_1-\lambda_2$, $\mu=\lambda_3-\lambda_4$,
$M = \lambda_1 - \lambda_2 - \lambda_3 + \lambda_4$, and $\theta$
($\varphi$) is the polar (azimuthal) scattering angle. $d$ is the
usual Wigner rotation matrix. This expression holds for gauge bosons
with $\lambda = 0,\pm1$, and for scalars (Higgs or ALP) with
$\lambda\equiv 0$; the fermion case will be addressed below. For
further details and conventions see
Ref.~\cite{Corbett:2014ora}. \smallskip

In the limit $S \gg (M_{V_1}+M_{V_2})^2$, partial-wave unitarity for a given \emph{elastic} channel requires that
\begin{equation}
  |T^J
  ({V_1}_{\lambda_1}{V_2 }_{\lambda_2} \to {V_1}_{\lambda_1}{V_2}_{\lambda_2}) 
| \le 1 \,.
\label{eq:unitcond}
\end{equation}
The most stringent bounds are obtained by diagonalizing $T^J$ in the
particle and helicity space and then applying the condition in
Eq.~(\ref{eq:unitcond}) to each of the eigenvalues. This is the
approach which we follow.  \smallskip

We start by calculating the scattering amplitudes for all possible
combinations of bosons and helicities generated by the SM extended
with the Lagrangian in Eq.~\eqref{eq:L_alp} for a given total electric
charge $Q=2, 1, 0$ and that give non-vanishing projections on a given
partial wave $J$ proportional to some ALP coupling.  Conservation of
color implies that initial or final states with color have to be
considered independently of those in a color singlet state.  So one is
led to consider separately the $T^J$ ($\overline{T}^J$) amplitude
matrices for processes with color singlet (octet) in the initial and
final states.  One must also take into account that parity
conservation at tree level implies the relation
\begin{equation}
 T^J({V_1}_{\lambda_1}{V_2 }_{\lambda_2} \to {V_3}_{\lambda_3}{V_4}_{\lambda_4})
=(-1)^{\lambda_1-\lambda_2-\lambda_3+\lambda_4}
  T^J({V_1}_{-\lambda_1}{V_2 }_{-\lambda_2} \to {V_3}_{-\lambda_3}{V_4}_{-\lambda_4}) 
\;\;.
\end{equation}
and leads to a reduction of the number of independent helicity
amplitudes.  Time-reversal invariance further reduces the number of
helicity amplitudes that need to be evaluated. \smallskip

Altogether, the initial/final states contributing a priori to the $T^J$ matrices
for each value of $Q$ and $J$ are:
\begin{equation}
\footnotesize
\hspace{-5mm}
  \begin{array}{c |ccccccccccccccccl}
(Q,J)			&
{\rm States}	&&&&&&&&&&&&&&&& 
\rm{Total} \\
\hline
\hline
(2,0) 		&	W^+_\pm W^+_\pm		& W^+_0W^+_0
&&&&&&&&&&&&&&& 3\\[+0.1cm]
\hline
(2,1) 		&      W^+_\pm W^+_0 	        &W^+_0W^+_\pm
&&&&&&&&&&&&&&& 4	\\[+0.1cm]
\hline
(1,0) 		&	W^+_\pm Z_\pm		&W^+_0Z_0		
&W^+_\pm \gamma_\pm		& W^+_0 a
&&&&&&&&&&&&&   6
\\[+0.1cm]
\hline
(1,1) 		&	{W^+_0Z_0}		&{W^+_\pm Z_0}	&
{W^+_0Z_\pm} &W^+_\pm Z_\pm	&	{W^+_0\gamma_\pm}
&W^+_\pm\gamma_\pm   &  {W^+_\pm a} & {W^+_0 a} & W^+_\pm H
&&&&&&&&      16\,  \\ [+0.1cm]
\hline
(0,0) 		&	W^+_\pm W^-_\pm		&W^+_0W^-_0		
		&Z_\pm Z_\pm	&Z_0Z_0	  	&	Z_\pm\gamma_\pm		
&\gamma_\pm\gamma_\pm &Z_0H	& 	G_\pm G_\pm 
& Z_0 a & H a& aa&  HH 
&&&& & 17\,\\[+0.1cm]
\hline
(0,1)		     &{W^+_0W^-_0}			&{ W^+_\pm W^-_0}
& {W^+_0 W^-_\pm}	&W^+_\pm W^-_\pm	&	Z_\pm Z_0
& Z_0Z_\pm 		 &Z_\pm\gamma_\pm	    & Z_0\gamma_\pm	
& Z_0H			 &Z_\pm H	&\gamma_\pm H	
&  {Z_0 a} &  {Z_\pm a} &  {\gamma_\pm a} & H a& &
26
\\[+0.1cm]
\end{array}
\label{eq:TJstates}
\end{equation}
and correspondingly the states contributing to the $\overline{T}^J$ matrices are:
\begin{equation}
\begin{array}{c |cccccc l}
(Q,J)			&{\rm States}&&&&& \rm{Total} \\
\hline
\hline
(1,0) 		&	{W^+_\pm G_\pm} &&&&&
2
\\[+0.1cm]
\hline
(1,1) 		&	{W^+_0 G_\pm} & {W^+_\pm G_\pm} &&&&
4\\ [+0.1cm]
\hline
(0,0)           &  {Z_\pm G_\pm} & 	 {\gamma_\pm G_\pm} &&&&
4 \\[+0.1cm]
\hline
(0,1)		& G_\pm G_\pm & {G_\pm a}
& {Z_\pm G_\pm}& {Z_0 G_\pm}& {\gamma_\pm G_\pm}  
& 10
\\[+0.1cm]
\end{array}
\label{eq:TbJstates}
\end{equation}
where upper indices indicate charge and lower indices helicity.  We
also display in Eqs.~(\ref{eq:TJstates}) and (\ref{eq:TbJstates}) the
dimensionality of the particle and helicity matrix for each
independent $(Q \, , J)$ channel.  In Eq.~\eqref{eq:TJstates} the states $HH$ 
and $W^\pm H$ are only present when the dimension 6 operator is considered.

\smallskip

We list in Tables~\ref{tab:ampli1}--~\ref{tab:ampli4}, that are shown
in the Appendix, 
the expressions for the most $S$-divergent part of the amplitudes for
the channels which give the dominant contribution to the $T^J$ and
$\overline{T}^J$ matrices.

\smallskip

{\bf Bounds on individual operators.}  As seen in
Tables~\ref{tab:ampli1}--~\ref{tab:ampli3}, for processes with zero or
two ALPs as external states, the most energy-divergent amplitudes
occur for scattering of transversely polarized gauge bosons, as
expected.  These amplitudes are all proportional to the product of two
axion couplings, therefore, the two powers of their momentum involved
in the coupling of ALP to the gauge boson generate the leading
$S/f_a^2$ dependence. A good fraction of them contributes to $J=0$
matrices, which are, a priori, expected to lead the strongest
bounds. Furthermore, for amplitudes with gluon pairs,
the strongest bounds are obtained for
the gluon pair in the singlet color state
$ \frac{1}{N_C^2-1}\displaystyle \sum_{a=1}^{N_C^2-1} \ket{G^a G^a}$.

\smallskip

Altogether from the diagonalization of the $J=0$ matrices and assuming
only one non-zero coupling at a time we find that the largest
eigenvalues correspond to the $Q=0$, $T^0$ matrix and read
\begin{align}
&\frac{1+\sqrt{97}}{16{\pi}} \, \frac{S}{f_a^2}\, \CW^2\,,
&
&\frac{1+\sqrt{33}}{16{\pi}} \, \frac{S}{f_a^2}\, \CB^2\,,
&
&\frac{4\,(N_C^2-1)}{{\pi}} \, \frac{S}{f_a^2}\, \CG^2\,
&
&{\rm and}\qquad\frac{1}{32\pi}\frac{S}{f_a^2}\CaPhitwo\,,
\label{eq:eigenvalues}
\end{align}
respectively. Applying the condition in Eq.~(\ref{eq:unitcond}) to
each of these eigenvalues we obtain the bounds
\begin{eqnarray}
 |\CW|&\leq& {2.1}\, \frac{f_a}{\rm TeV} \, \left(\frac{\rm TeV}{\sqrt{S}}\right)\, ,
\label{eq:boundcw1}  \\
 |\CB|&\leq& {2.7}\, \frac{f_a}{\rm TeV} \, \left(\frac{\rm TeV}{\sqrt{S}}\right)\, ,
\label{eq:boundcb}  \\
 |\CG|&\leq& {0.31}\, \frac{f_a}{\rm TeV} \, \left(\frac{\rm TeV}{\sqrt{S}}\right)\, ,
 \label{eq:boundcg}  \\
|\CaPhitwo|&\leq& 101\, \frac{f_a^2}{\unit{TeV^2}} \, \left(\frac{\rm TeV^2}{S}\right) \,.
\label{eq:boundcaphi2}
 \end{eqnarray}

We observe that the constraint on the dimension 6 operator $O_{a\Phi}^{(2)}$
is dominated by scattering amplitudes with 2 ALP external states.

Unlike the amplitudes with even number of ALP in the external states,
some helicity amplitudes with only one ALP in either the initial or
final state have a leading behavior $S^{\frac{3}{2}}/(f_a\,M_W^2)$, as
seen in Table~\ref{tab:ampli4} \footnote{This is at
  variance with what is found for effective interactions in the
  SMEFT \cite{Corbett:2014ora,Corbett:2017qgl,Almeida:2020ylr} for
  which all operators of a given dimension lead to most divergent
  amplitudes with the same power of $S$, that is, $S$ for
  dimension-six operators and $S^2$ for dimension-8 operators.}. These
amplitudes involve two longitudinally polarized gauge bosons, whose
polarization vectors are proportional to $\sqrt{S}$, and one transversely
polarized gauge boson whose momentum contributes another power of
$\sqrt{S}$.  This configuration can only be generated by a combination
of the SM vertices and those induced by $O_{\tilde W}$ and,
consequently, the amplitudes involve a single power of the $\CW$
coupling and of the SM coupling $e$, and do not depend on any other
Wilson coefficient.  As seen from the scattering angle dependence of
the amplitudes in Table~\ref{tab:ampli4}, they contribute only to the
$T^{J=1}$ matrix with either $Q=0$ or $Q=1$.  Diagonalizing these we
find that the largest eigenvalue is
\begin{equation}
\frac{\sqrt{1+c_w^2+2\,c_w^4}}{24 \, \pi} \,\frac{S^{3/2}}{f_a\, M_W^2}
\, e\, \CW\;,
\end{equation}
where $c_w$ is the cosine of the weak mixing angle. Therefore, the
condition in Eq.~\eqref{eq:unitcond} implies the constraint on $\CW$
\begin{equation}
  |\CW| \leq 0.14 \,
    \frac{f_a}{\rm TeV}  \left(\frac{\rm TeV}{\sqrt{S}}\right)^3 \; .
\label{eq:boundcw}
\end{equation}

Comparing the bounds on $\CW$ from $J=0$-wave unitarity,
Eq.~\eqref{eq:boundcw1}, and from $J=1$-wave unitarity,
Eq.~\eqref{eq:boundcw}, we find that the constraint derived from the
$J=1$ amplitudes is the strongest for
$$\sqrt{S}> 260 \;{\rm GeV}\;.$$

\smallskip {\bf Including multiple operators simultaneously.}  Fixing
$\CaPhitwo=0$ (or equivalently barring cancellations between dimension
5 and 6 terms) and allowing multiple dimension 5 operators to vary
simultaneously does not alter significantly the bounds reported
above. For $\CW$ this is obvious, because the leading constraint
Eq.~\eqref{eq:boundcw} is genuinely independent of the other Wilson
coefficients. For $\CB$ and $\CG$ this can be understood considering
that $\CG$ is dominantly constrained by
$G_\pm G_\pm\to G_\pm G_\pm$ scattering in the color singlet
  channel, which is independent of $\CB$.  We have also verified this
explicitly by diagonalizing the $Q=0$ $T^{J=0}$ matrix with $\CB$,
$\CG$ present at the same time. The diagonalization can still be done
analytically though the resulting expressions for the eigenvalues are
not particularly illuminating. Imposing the unitarity limits on those
eigenvalues yields the same bounds as in Eqs.~\eqref{eq:boundcb}
and~\eqref{eq:boundcg}.

\smallskip

Allowing all operators of dimension 5 and 6 to be present
simultaneously (i.e. allowing cancellations between both orders) we
find that the largest eigenvalues are
\begin{align}
&\frac{5}{8{\pi}} \, \frac{S}{f_a^2}\, \CW^2\,,
&
&\frac{1}{8{\pi}} \, \frac{S}{f_a^2}\, \CB^2\,,
&
&\frac{4\,(N_C^2-1)}{{\pi}} \, \frac{S}{f_a^2}\, \CG^2\,
&{\rm and}\qquad\qquad\frac{1}{32\pi}\frac{S}{f_a^2}\CaPhitwo\,,
\label{eq:eigenvalues2}
\end{align}
and correspondingly the unitarity limits on the Wilson coefficients are:
\begin{eqnarray}
 |\CW|&\leq& {2.2}\, \frac{f_a}{\rm TeV} \, \left(\frac{\rm TeV}{\sqrt{S}}\right)\, ,
\label{eq:boundcw2}  \\
 |\CB|&\leq& {5.0}\, \frac{f_a}{\rm TeV} \, \left(\frac{\rm TeV}{\sqrt{S}}\right)\, ,
\label{eq:boundcb2}  \\
 |\CG|&\leq& {0.31}\, \frac{f_a}{\rm TeV} \, \left(\frac{\rm TeV}{\sqrt{S}}\right)\, ,\label{eq:boundcg2} \\
 |\CaPhitwo|&\leq& 101\, \frac{f_a^2}{\unit{TeV^2}} \, \left(\frac{\rm TeV^2}{S}\right) \,.
\label{eq:boundcaphi22}
\end{eqnarray}
These results hold irrespective of whether $\CW,\,\CB$ and $\CG$ are
included simultaneously or individually.  It is also worth noting that
the bounds on $\CG$ and $\CaPhitwo$ are unchanged compared to the
individual limits~\eqref{eq:boundcg} and
\eqref{eq:boundcaphi2}. Considering that $\CW$ is always dominantly
constrained by Eq.~\eqref{eq:boundcw}, we conclude that only the
unitarity constraints on $\CB$ depends significantly on whether
$\CaPhitwo$ is included or not.

\smallskip

{\bf Truncating at dimension-5.} Finally, it can be interesting to
investigate bounds on the dimension-5 interactions only.  As we have
seen above, the most stringent bounds on $\CW$ originates from
processes exhibiting just one dimension-5 vertex, therefore, it is not
modified when we truncate the EFT expansion to $\O(f_a^{-1})$.

\smallskip

In order to obtain limits on $\CB$ and $\CG$ independently of
assumptions about $\CaPhitwo$, we can restrict our analysis to a
subspace of initial states such that contributions of the dimension-6
operator are negligible for all the scattering amplitudes
retained. This is achieved by eliminating ``flavor'' states in
Eqs.~(\ref{eq:TJstates}) and (\ref{eq:TbJstates}) that lead to
processes containing two ALP external legs.  We can re-derive the
constraints on this flavor subspace and we obtain that the largest
eigenvalues come from the $Q=0$, $T^0$ matrix and they coincide with
those in Eq.~\eqref{eq:eigenvalues2}, leading to the bounds in
Eqs.~\eqref{eq:boundcw2}--\eqref{eq:boundcg2}.  The result is the same
irrespective of whether $\CB$ and $\CG$ are included simultaneously or
individually.

\smallskip

\subsection{Helicity amplitudes involving fermions}
\label{sec:fermions}

ALP couplings to fermions can contribute to processes
\begin{equation}
{f_1}_{\sigma_1}{\bar {f_2}}_{\sigma_2} \to 
{V_3}_{\lambda_3}{V_4}_{\lambda_4} \, ,
\label{eq:ffVV}
\end{equation}
which can also violate unitarity. In this case the partial-wave expansion is
given by
\begin{equation}
\mathcal{M}
({f_1}_{\sigma_1}{\bar {f_2}}_{\sigma_2} \to 
{V_3}_{\lambda_3}{V_4}_{\lambda_4})
 =16 \pi \sum_J
\left ( 2J+1 \right)~\delta_{\sigma_1, \sigma_2}
d_{\sigma_1-\sigma_2,\lambda_3-\lambda_4}^{J}(\theta)
~ T^J({f_1}_{\sigma_1}{\bar {f_2}}_{\sigma_2} \to 
{V_3}_{\lambda_3}{V_4}_{\lambda_4}) \,.
\label{eq:helamp2}
\end{equation}
In principle,
${f_1}_{\sigma_1}{\bar {f_2}}_{\sigma_2}\to
{V_3}_{\lambda_3}{V_4}_{\lambda_4}$ amplitudes of a given $J$ partial
wave can be incorporated together with the
${V_1}_{\lambda_1}{V_2 }_{\lambda_2} \to
{V_3}_{\lambda_3}{V_4}_{\lambda_4}$ amplitudes in the corresponding
$T^J$ matrix by extending the basis of states to incorporate the
relevant ${f_1}_{\sigma_1}{\bar {f_2}}_{\sigma_2}$ combinations
contributing to a given $Q$; see, for example,
Ref.~\cite{Chanowitz:1978uj}.  However, we find that the most energy
divergent amplitudes for fermion-antifermion scattering grow at most
as $\sqrt{S}$ and, therefore, the contributions from the ALP-fermion
couplings enter with different power of $S$ with respect to the
ALP-gauge-boson couplings in the eigenvalues of this generalized $T^J$
matrices.  Thus, in order to derive independent unitarity constraints
on the $C_{f\Phi}$ couplings we find it more convenient to follow the
alternative procedure presented in Ref.~\cite{Baur:1987mt} and relate
the corresponding
${f_1}_{\sigma_1}{\bar {f_2}}_{\sigma_2}\to
{V_3}_{\lambda_3}{V_4}_{\lambda_4}$ amplitude to that of the elastic
process
\begin{equation}
{f_1}_{\sigma_1}{\bar {f_2}}_{\sigma_2} \to 
{f_1}_{\sigma_1}{\bar {f_2}}_{\sigma_2}   \,.
\end{equation}
In this case the unitarity relation is
\begin{eqnarray}
{\rm Im}[T^J({f_1}_{\sigma_1}{\bar {f_2}}_{\sigma_2} 
\to {f_1}_{\sigma_1}{\bar {f_2}}_{\sigma_2})]&=&
\left|T^J({f_1}_{\sigma_1}{\bar {f_2}}_{\sigma_2}\to {f_1}_{\sigma_1}{\bar {f_2}}_{\sigma_2})\right|^2 \label{eq:unitff}\\ 
&&+\sum_{{V_3}_{\lambda_3},{V_4}_{\lambda_4}}
\left|T^J({f_1}_{\sigma_1}{\bar {f_2}}_{\sigma_2}\to 
{V_3}_{\lambda_3}{V_4}_{\lambda_4})\right|^2+
\sum_N
\left|T^J({f_1}_{\sigma_1}{\bar {f_2}}_{\sigma_2}\to N)\right|^2 \, ,\nonumber
\end{eqnarray}
where we take the limit
$S \gg (M_{V_3}+M_{V_4})^2\;,\;(M_{f_1}+M_{f_2})^2$.  $N$ represents
any state which ${f_1}_{\sigma_1}{\bar {f_2}}_{\sigma_2}$ can
annihilate into which does not consist of two
bosons. Eq.~(\ref{eq:unitff}) is a quadratic equation for
${\rm Im}[T^J({f_1}_{\sigma_1}{\bar {f_2}}_{\sigma_2} \to
{f_1}_{\sigma_1}{\bar {f_2}}_{\sigma_2})]$ which only admits a
solution if
\begin{equation}
\sum_{{V_3}_{\lambda_3},{V_4}_{\lambda_4}}
\left|T^J({f_1}_{\sigma_1}{\bar {f_2}}_{\sigma_2}\to 
{V_3}_{\lambda_3}{V_4}_{\lambda_4})\right|^2 \leq \frac{1}{4} \; . 
\label{eq:unitcond2}
\end{equation}
The strongest bounds can be found by considering some optimized linear
combinations
\begin{equation}
   \ket{X} = \sum_{f_1,\sigma_1} x_{f_2,\sigma_2} 
   \ket{{f_1}_{\sigma_1} {\bar{f_2}}_{\sigma_2} }
\end{equation}
with the normalization condition $\sum_{f\sigma} | x_{f\sigma}|^2=1$,
for which the amplitude $T^J(X\to {V_3}_{\lambda_3}{V_4}_{\lambda_4})$
is the largest. \smallskip

In this approach, processes of fermion scattering into one gauge boson
and one ALP provide independent constraints on the ALP-fermion
coupling.  As mentioned above, the most divergent relevant helicity
amplitudes grow as $\sqrt{S}$ and are listed in
Table~\ref{tab:fermions}.  For couplings to leptons of a given
generation the strongest bounds are obtained with
$\ket{X} =\frac{1}{\sqrt{2}} \ket{e^-_+\,e^+_+\,+\,{e^-_-\,e^+_- }}$
(or equivalently with $\ket{X} =\ket{{\nu_-\,e^+_-}}$).  For couplings
to quarks of a given generation, accounting for the $N_C=3$ color
states, the strongest bounds are obtained with
$\ket{X} =\dfrac{1}{\sqrt{2N_c}}\displaystyle\sum_{a=1}^{N_c} \ket{q^a_+\,\bar
  q^a_+\,+{\,q^a_-\,\bar q^a_-}}$, or equivalently with
$\dfrac{1}{\sqrt{N_c}}\displaystyle \sum_{a=1}^{N_c}\ket{{u^a_+\,\bar
    d^a_+}}$ and
$\dfrac{1}{\sqrt{N_c}}\displaystyle \sum_{a=1}^{N_c}\ket{u^a_-\,\bar
  d^a_-}$ ).  Furthermore, within the assumed flavour symmetry of the
axion coupling to fermions, the strongest bounds correspond to the
processes with fermions of the third generation and they read
\begin{alignat}{2}
  \left|C_{a\Phi}- C_{e\Phi}\right|&\leq \frac{16\,\pi}{|Y_\tau|}
  \left(\frac{f_a}{\sqrt{S}}\right)&&= \frac{50}{|Y_\tau|}
 \frac{f_a}{\rm TeV}  \left(\frac{\rm TeV}{\sqrt{S}}\right) 
  \;, \\
  \left|C_{a\Phi}+ C_{u\Phi}\right|&\leq \frac{16\,\pi }{\sqrt{N_C} \,|Y_t|}
  \left(\frac{f_a}{\sqrt{S}}\right)&&=
\frac{29 }{|Y_t|}\frac{f_a}{\rm TeV}  \left(\frac{\rm
    TeV}{\sqrt{S}}\right)  \;,
\label{eq:cubound}
\\
  \left|C_{a\Phi}- C_{d\Phi}\right|&\leq \frac{16\,\pi}{\sqrt{N_C} \, |Y_b|}
  \left(\frac{f_a}{\sqrt{S}}\right)&&=
\frac{29 }{|Y_b|}\frac{f_a}{\rm TeV}  \left(\frac{\rm TeV}{\sqrt{S}}\right)    \;.  
 \end{alignat}
Because these bounds are inversely proportional to the Yukawa
coupling of the fermion and involve larger coefficients, we conclude
that the unitarity constraints on the ALP-fermion couplings are
orders of magnitude weaker than those on ALP-gauge-boson couplings
even for the coupling to the up-quarks. Moreover, the operator
${\cal O}_{a\Phi}$ should only be considered in a scenario where the
fermionic operators are absent. In this case, the most stringent
unitarity constraint on its coupling originates from
Eq.~\eqref{eq:cubound}.

\smallskip

\begin{figure}[t]\centering
\includegraphics[width=0.6\textwidth]{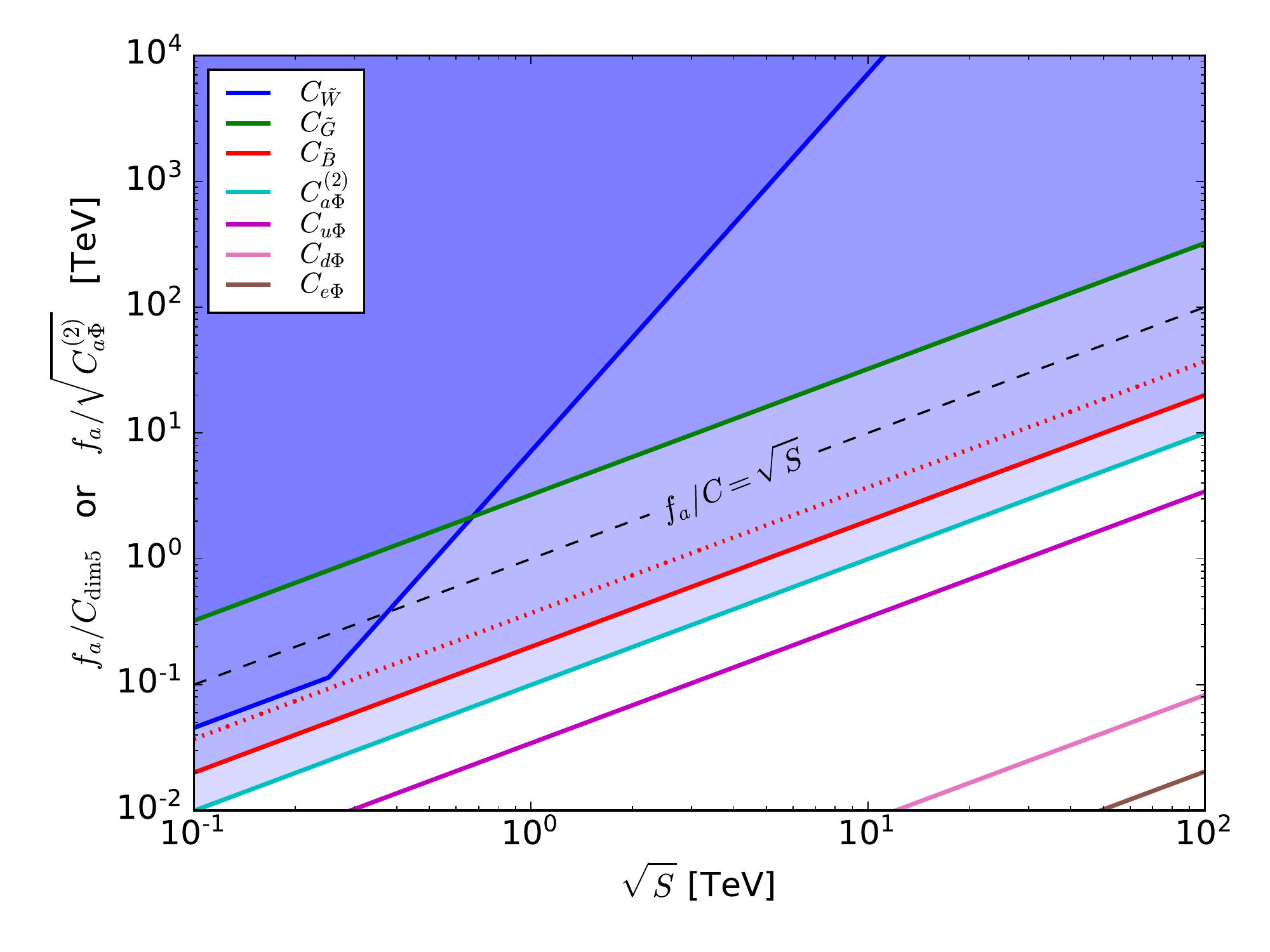}
\caption{Summary of unitarity bounds derived in this work. The shaded
  regions indicate allowed values of $f_a/C$ ($f_a/\sqrt{C}$ in the
  case of $\CaPhitwo$) and $\sqrt{S}$ for each bosonic interaction.
  For $\CW$ we plot the most stringent bound between
  Eq.~\eqref{eq:boundcw1} and~\eqref{eq:boundcw}. The solid red line
  indicates the bound on $\CB$ in Eq.\eqref{eq:boundcb2}, derived
  allowing $\CaPhitwo$ to vary simultaneously. The individual bound on
  $\CB$ given in  Eq.~\eqref{eq:boundcb} is drawn as a dotted line in the same
  color.  The bounds on fermionic operators are always subdominant and
  are only marked as solid lines. The bound on $C_{a\Phi}$, that
  should only considered in a setup where the fermionic operators are
  absent, coincides with that on $C_{u\Phi}$.  Finally, the grey
  dashed line marks the diagonal for reference. }
\label{fig:bounds}
\end{figure}

\section{Conclusions}
\label{sec:conclu}

We have derived maximal constraints on the effective interactions of
Axion-Like-Particles from partial-wave unitarity in $2\to2$ scattering
processes. Our results are summarized in Fig.~\ref{fig:bounds}. They
hold in the kinematic regime where $\sqrt{S} \gg v$ and the ALP mass
was also implicitly taken to be $m_a\ll\sqrt{S}$. Furthermore the
consistency of the ALP EFT expansion requires $\sqrt{S}\ll f_a $.

\smallskip

We find that, for fixed $C/f_a$, the most stringent unitarity bound is
imposed on $\CW$ in $VV\to Va$ scattering processes, while the weakest
limits are on ALP-fermion interactions. The constraints exhibit only a
limited dependence on whether the effective operators are taken
individually or allowed to vary simultaneously, signaling that each of
them is dominantly constrained in a class of scattering amplitudes
that is nearly orthogonal to the others.

\smallskip

The constraints we have derived can be particularly relevant for ALP
searches at colliders~\cite{Jaeckel:2012yz, Mimasu:2014nea,
  Jaeckel:2015jla, Brivio:2017ije, Bauer:2017ris, Bauer:2018uxu,
  Frugiuele:2018coc, Craig:2018kne, Ebadi:2019gij,Yue:2019gbh,
  Gavela:2019cmq, Inan:2020aal, Haghighat:2020nuh, Goncalves:2020bqi,
  Florez:2021zoo, VBS} where, depending on their masses, ALP particles
could be observed in $a\to XX$ decays, with $X$ a generic SM state, in
$Xa$ associated production (with the ALP either going undetected or
decaying to photons or light fermions) or in non-resonant $2\to2$
scattering processes, where the ALP appears as an off-shell internal
line~\cite{Gavela:2019cmq, Florez:2021zoo, VBS}.

\smallskip 
In this respect, let us stress that our results should not be
interpreted as strict unitarity constraints on any specific process
used in the ALP searches, in the sense that it might be difficult to directly identify the kinematic
information available 
with the subprocess center-of-mass energy of an individual $2\to2$
scattering.
Notwithstanding, unitarity bounds must be satisfied in the event
generation and, consequently, can affect the shapes of expected
distributions used in the searches.

\smallskip

For example, recently, the ATLAS collaboration has searched for axions
in events with an energetic jet~\cite{Aad:2021egl} or a
photon~\cite{Aad:2020arf} and missing transverse momentum.  The
monojet analysis~\cite{Aad:2021egl} constrains the axion coupling to
gluons to satisfy $\CG/f_a<0.008$ TeV$^{-1}$ at 95\% CL. Using
Eq.~\eqref{eq:boundcg2}, we find that for the largest allowed coupling
in this search unitarity is preserved up to center mass-of-mass energy
of 39 TeV, clearly beyond the LHC reach.  On the other hand, the
mono-photon analysis limits the $\CW$ coupling to satisfy
$\CW/f_a<0.12$ TeV$^{-1}$ at 95\% CL.  From Eq.~\eqref{eq:boundcw} we
read that for $\CW/f_a$ at the 95\% CL boundary, unitarity is
violated in subprocesses with center-of-mass energy greater than 1.04
TeV. We conclude that the tail of the expected missing $E_T$
distribution should be analyzed cautiously and the unitarity
constraints could have an impact in the derivation of the mono-photon
search bound.

The unitarity bounds derived in this work would be also relevant in the 
event that an ALP signal will be detected in the future (independently 
of the energy regime at which the experimental search is conducted), 
leading to a defined measurement of one or more ALP couplings. 
In this case, unitarity bounds would provide an upper limit to the mass 
scale of the new physics sector the ALP originates from and motivate 
further searches in this energy region.

\smallskip



\section*{Acknowledgments}

O.J.P.E. is supported in part by Conselho Nacional de Desenvolvimento
Cent\'{\i}fico e Tecnol\'ogico (CNPq) and by Funda\c{c}\~ao de Amparo
\`a Pesquisa do Estado de S\~ao Paulo (FAPESP) grant 2019/04837-9.
M.C.G-G is supported by spanish grant PID2019-105614GB-C21, by USA-NSF
grant PHY-1915093, and by AGAUR (Generalitat de Catalunya) grant
2017-SGR-929. The authors acknowledge the support of European ITN
grant H2020-MSCA-ITN-2019//860881-HIDDeN.


\clearpage
\section*{Helicity amplitudes at leading order in \tpdf{$S$}{S}}\label{app:tables}
We present here the list of unitarity violating amplitudes for
all the $2\rightarrow 2$ scattering processes considered in the evaluation
of the unitarity constraints. 

\begin{table}[h]\centering
\renewcommand{\arraystretch}{1.3}
$\begin{array}{|c|cccc|c|}
\hline
    & \lambda_1 &\lambda_2 &\lambda_3 &\lambda_4 
& {\cal M}\left(\times \frac{S}{f_a^2}\right)\\\hline
    W^+W^+\rightarrow W^+W^+ 
    &  1 &-1 &  1 &  1 &  2 \,\CW^2\\
    & -1 & 1 &  \mp 1 &  \pm1 &  X_\pm \,\CW^2\\    
\hline
W^+ Z\rightarrow W^+ Z
& -1 &  \mp 1 &  1 &  \pm 1 &  
\pm 2 \,\CW \;{\rm X_-} \left(\,s_w^2\,\CB \,+\,c_w^2 \,\CW\right) \\
\hline
W^+ Z\rightarrow W^+ \gamma
& -1 &  \mp 1 &  1 &  \pm 1 &  
 \mp 2 \,c_w \,\CW \, s_w \;{\rm X_-} (\,\CB-\,\CW) \\
 \hline
 W^+ \gamma\rightarrow W^+ \gamma
& -1 &  \mp 1 &  1 &  \pm 1 &  
\pm 2 \,\CW \;{\rm X_-}
\left(\,c_w^2\,\CB \,+\,s_w^2 \,\CW\right) \\
\hline
W^+ W^-\rightarrow W^+ W^-
& -1 &  -1 &  -1 &  -1 &  4 \,\CW^2\\
& -1 &  \mp 1 &  1 &  \pm 1 &  -2 \,\CW^2 \;{\rm X_\pm}\\
\hline
W^+ W^-\rightarrow Z  Z 
& -1 &  -1 &  \mp 1 &  \mp 1 &  \pm 2 \sqrt{2} \,\CW
\left( \,s_w^2\, \CB +\,c_w^2 \,\CW\right)\\
\hline
W^+ W^-\rightarrow Z \gamma
& -1 &  -1 &  \pm 1 &  \pm 1 &  
 \pm 4 \,c_w \,\CW \, s_w (\,\CB-\,\CW)\\
\hline
W^+ W^-\rightarrow \gamma \gamma 
& -1 &  -1 &  \mp 1 &  \mp 1 &  
\pm 2 \sqrt{2} \,\CW
\left(\,c_w^2\,\CB \,+\,s_w^2 \,\CW\right)\\
\hline
Z Z \rightarrow Z Z
& -1 &  -1 &  -1 &  -1 &  
 2 \left(\,c_w^2 \,\CW-\, s_w^2\,\CB \right)^2\\
& -1 &  1 &  \mp 1 &  \pm 1 &  
 -X_\pm\, \left(\,c_w^2 \,\CW-\, s_w^2\,\CB \right)^2\\
 \hline
Z Z \rightarrow Z \gamma
& -1 &  -1 &  -1 &  -1 &  
 2 \sqrt{2} \,c_w \, s_w (\,\CB-\,\CW) \
\left(\, s_w^2\,\CB -\,c_w^2 \,\CW\right)\\
& -1 &  -1 &  \mp 1 &  \pm1 &  
-\sqrt{2} \,X_\pm\, c_w \, s_w (\,\CB-\,\CW) \
\left(\, s_w^2\,\CB -\,c_w^2 \,\CW\right)\\
\hline
Z Z \rightarrow \gamma \gamma
& -1 &  -1 &  -1 &  -1 &  
-2 \left(\,(\CB^2+\CW^2)\,c_w^2 \, s_w^2-\,\CB \,\CW \
\left(\,c_w^4+\, s_w^4\right)\right)\\
& -1 &  -1 &  1 &  1 &  
- 2 \,\CB \,\CW \\
& -1 &  1 &  \mp 1 &  \pm 1 &  
X_\pm\, (\CB-\CW)^2\,c_w^2 \, s_w^2 \\
\hline
Z \gamma \rightarrow Z \gamma
& -1 &  -1 &  -1 &  -1 &  4 \,c_w^2 \, s_w^2 (\,\CB-\,\CW)^2\\
& -1 &  -1 &  1 &  1 &  -2 \,\CB \,\CW \;{\rm X_-} \\
& -1 &  1 &  \mp 1 &  \pm 1 &  
-2 \,X_\pm\, (\CB-\CW)^2\,c_w^2 \, s_w^2 \\
\hline
\gamma \gamma \rightarrow \gamma \gamma
& -1 &  -1 &  -1 &  -1 &  
2 \left(\,c_w^2\,\CB \,+\,s_w^2 \,\CW\right)^2
\\
& -1 &  1 &  \mp 1 &  \pm 1 &  
- X_\pm\, \left(\,c_w^2\,\CB \,+\,s_w^2 \,\CW\right)^2
\\
 \hline
W^+ W^- \rightarrow G^a G^b
& -1 &  -1 &  \mp 1 &  \mp 1 &  
\pm  8 \sqrt{2} \,\CG \,\CW \
\,\delta_{ab}\\
\hline
Z Z \rightarrow G^a G^b
& -1 &  -1 &  \mp 1 &  \mp 1 &  
\pm 8 \,\CG
\left(\,s_w^2\,\CB+\,c_w^2 \,\CW\right) \,\delta_{ab}\\
\hline
Z \gamma \rightarrow G^a G^b
& -1 &  -1 &  \mp 1 &  \mp 1 &  
 \mp 8 \sqrt{2} \,\CG \,c_w \, s_w (\,\CB-\,\CW) \
\,\delta_{ab}\\
 \hline
 \gamma \gamma \rightarrow G^a G^b
& -1 &  -1 &  \mp 1 &  \mp 1 &  
\pm 8 \,\CG \left(\,s_w^2\,\CB \,+\,c_w^2 \,\CW\right) \,\delta_{ab}\\
\hline
G^a G^b \rightarrow G^c G^d
& -1 &  -1 &  -1 &  -1 &  
 32 \,\CG^2 \, \delta_{ab}\, \delta_{cd} \\
 & -1 &  -1 &  1 &  1 &  
 16 \,\CG^2 (\;{\rm X_-} \,\delta_{ac}\, \delta_{bd}+\;{\rm X_+} \, \delta_{ad}\, \delta_{bc}-
 2 \, \delta_{ab}\, \delta_{cd})\\
 & -1 &  1 &  -1 &  1 &  
 -4\,X_+ \,\CG^2 \, \delta_{ac}\, \delta_{bd} \\
  & -1 &  1 &  1 &  -1 &  
 -4\,X_- \,\CG^2 \, \delta_{ad}\, \delta_{bc} \\
 \hline
\end{array}$
  \caption{
  Leading contributions to the helicity amplitudes for the channels
  involving SM bosons and even number of gluons in the
  initial or final state and with projections in  $J=0$.
  They contribute to the $T^J$ matrices with $Q=2,1,0$ and $J=0,1$ 
  In the expressions ${\rm X_+}=(1+\cos x)$ and
  ${\rm X_-}=(1-\cos x)$ where $x$ is the polar angle.}
\label{tab:ampli1}
\end{table}

\begin{table}[h]\centering
\renewcommand{\arraystretch}{1.5}
$\begin{array}{|c|cccc|c|}
\hline
    & \lambda_1 &\lambda_2 &\lambda_3 &\lambda_4 
& {\cal M}\left(\times \frac{S}{f_a^2}\right)\\\hline
W^+ a  \rightarrow W^+ a
& -1 &  0 &  -1 &  0 &  -2 \,\CW^2 \;{\rm X_-}\\
& -1 &  0 &  1 &  0 &  4 \,\CW^2 \;{\rm X_-}\\
& 0 & 0 & 0 & 0 & \frac{1}{2} \CaPhitwo\; {\rm X_-}\\
\hline
W^+ W^-  \rightarrow a a
& -1 &  -1 &  0 &  0 &  4 \sqrt{2} \,\CW^2\\
& -1 &   1 &  0 &  0 &  -2 \sqrt{2} \,\CW^2\\
& 0 & 0 & 0 & 0 & \frac{1}{\sqrt{2}}\,\CaPhitwo\\
\hline
Z Z  \rightarrow a a
& -1 &  -1 &  0 &  0 &  
4 \left(\,s_w^2\,\CB^2 \,+\,c_w^2 \,\CW^2\right)\\
& -1 &  1 &  0 &  0 &  
-2 \left(\,s_w^2\,\CB^2 \,+\,c_w^2 \,\CW^2\right)\\
& 0 & 0 & 0 & 0 & \CaPhitwo\\
\hline
Z \gamma  \rightarrow a a
& -1 &  -1 &  0 &  0 &  
 -4 \sqrt{2} \,c_w \, s_w \left(\CB^2-\,\CW^2\right)\\
& -1 &  1 &  0 &  0 &  
 2 \sqrt{2} \,c_w \, s_w \left(\CB^2-\,\CW^2\right)\\
\hline
Z a  \rightarrow Z a
& -1 &  0 &  -1 &  0 &  
 -2 \;{\rm X_-} \left(\,s_w^2\,\CB^2 \,+\,c_w^2 \,\CW^2\right)\\
& -1 &  0 &  1 &  0 &  
 4 \;{\rm X_-} \left(\,s_w^2\,\CB^2 \,+\,c_w^2 \,\CW^2\right)\\
 & 0 & 0 & 0 & 0 & \frac{1}{2} \CaPhitwo\; {\rm X_-}\\
\hline
Z a  \rightarrow \gamma a
& -1 &  0 &  -1 &  0 &  
 2 \,c_w \, s_w \;{\rm X_-} \left(\CB^2-\,\CW^2\right)\\
& -1 &  0 &  1 &  0 &  
 -4 \,c_w \, s_w \;{\rm X_-} \left(\CB^2-\,\CW^2\right)\\
\hline
\gamma \gamma  \rightarrow a a
& -1 &  -1 &  0 &  0 &  
 4 \left(\,c_w^2\,\CB^2 +\,s_w^2\,\CW^2 \right)\\
& -1 &  1 &  0 &  0 &  
 -2 \left(\,c_w^2\,\CB^2 +\,s_w^2\,\CW^2 \right)\\
\hline
\gamma a  \rightarrow \gamma a
& -1 &  0 &  -1 &  0 &  
 - 2  {\rm X_-}\left(\,c_w^2\,\CB^2 +\,s_w^2\,\CW^2 \right)\\
& -1 &  0 &  1 &  0 &  
 4 {\rm X_-}
 \left(\,c_w^2\,\CB^2 +\,s_w^2\,\CW^2 \right) \\
\hline
G^a G^b  \rightarrow a a
& -1 &  -1 &  0 &  0 &  
16 \,\CG^2 \,\delta_{ab}\\
& -1 &  11 &  0 &  0 &  
-8 \,\CG^2 \,\delta_{ab}\\
\hline
H a  \rightarrow H a
& 0 & 0 & 0 & 0 & \frac{1}{2} \CaPhitwo\; {\rm X_-}\\
\hline
H H  \rightarrow a a
& 0 & 0 & 0 & 0 & \frac{1}{2} \CaPhitwo\; \\
\hline
  \end{array}$
  \caption{
  Leading contributions to the helicity amplitudes for the channels
  involving SM bosons and two ALPs and even number of gluons in the
  initial or final state.
  They contribute to the $T^J$ matrices with $Q=1,0$ and
  $J=0$ or $J=1$.
  In the expressions ${\rm X_+}=(1+\cos x)$ and
  ${\rm X_-}=(1-\cos x)$ where $x$ is the polar angle.}
  \label{tab:ampli2}
\end{table}

\begin{table}[h]\centering
\renewcommand{\arraystretch}{1.5}
$\begin{array}{|c|cccc|c|}
\hline
& \lambda_1 &\lambda_2 &\lambda_3 &\lambda_4 
& {\cal M}\left(\times \frac{S}{f_a^2}\right)\\\hline
 W^+ G^a\rightarrow W^+ G^b
& -1 &  \mp 1 &  1 &  \pm 1 &  
 \pm 8 \,\CG \,\CW \;{\rm X_-} \,\delta_{ab}\\
\hline
Z G^a \rightarrow Z G^b
& -1 &  \mp 1 &  1 &  \pm 1 &  
\pm 8 \,\CG \;{\rm X_-}
\left(\,s_w^2\,\CB +\,c_w^2 \,\CW\right) \,\delta_{ab}\\
\hline
Z  G^a \rightarrow \gamma G^b
& -1 &  \mp 1 &  1 &  \pm 1 &  
 \mp 8 \,\CG \,c_w \, s_w \;{\rm X_-} (\,\CB-\,\CW) \,\delta_{ab}\\
\hline
\gamma G^a \rightarrow \gamma G^b
& -1 &  \mp 1 &  1 &  \pm 1 &  
\pm 8 \,\CG \;{\rm X_-}
\left(\,c_w^2\,\CB \,+\,s_w^2 \,\CW\right) \,\delta_{ab}\\
\hline
G^a a  \rightarrow G^b  a
& -1 &  0 &  -1 &  0 &  
 -8 \,\CG^2 \;{\rm X_-} \,\delta_{ab}\\
& -1 &  0 &  1 &  0 &  
 16 \,\CG^2 \;{\rm X_-} \,\delta_{ab}\\
\hline
\end{array}$
  \caption{
  Leading contributions to the helicity amplitudes for the channels
 with one gluon in the initial and final state. 
   They contribute $\overline{T}^J$ with $Q=1,0$ and $J=0,1$.
   ${\rm X_+}=(1+\cos x)$ and
  ${\rm X_-}=(1-\cos x)$ where $x$ is the polar angle.}   
 \label{tab:ampli3}
 \end{table}

\begin{table}[h]\centering
\renewcommand{\arraystretch}{1.5}
$\begin{array}{|c|cccc|c|}
\hline
    & \lambda_1 &\lambda_2 &\lambda_3 &\lambda_4 
& {\cal M}\left(\times e\,\frac{S^{3/2}}{f_a M_W^2}\right)\\\hline    
 W^+ W^- \rightarrow Z a
& -1 &  0 &  0 &  0 &  \sqrt{2}\, \frac{c_w^2}{s_w} \,\CW \, {\rm Y}\\
& 0 &  -1 &  0 &  0 &  -\sqrt{2}\, \frac{c_w^2}{s_w} \,\CW \, {\rm Y}\\
& 0 &  0 &  -1 &  0 &  \sqrt{2}\, \frac{c_w}{s_w} \,\CW \, {\rm Y}
 \\
 \hline
 W^+ W^- \rightarrow \gamma a
& 0 &  0 &  -1 &  0 &  \sqrt{2} \,\CW \, {\rm Y}\\
\hline
W^+ Z \rightarrow W^+ a
& -1 &  0 &  0 &  0 &  -\sqrt{2}\, \frac{c_w^2}{s_w} \,\CW \, {\rm Y}
\\
& 0 &  -1 &  0 &  0 &  \sqrt{2}\, \frac{c_w^2}{s_w} \,\CW \, {\rm Y}\\
& 0 &  0 &  -1 &  0 &  -\sqrt{2}\, \frac{c_w^2}{s_w} \,\CW \, {\rm Y}\\
\hline
W^+ \gamma \rightarrow W^+ a
& 0 &  -1 &  0 &  0 &  \sqrt{2} \,\CW \, {\rm Y}\\
\hline
  \end{array}$
  \caption{
  Leading contributions to the helicity amplitudes for the channels
  involving SM bosons and one ALP.
  They contribute to the $T^J$ matrices with $Q=1,0$ and $J=1$ 
  In these amplitudes ${\rm Y}=\sin x$ where $x$ is the polar angle.}
\label{tab:ampli4}
\end{table}

\begin{table}[h]\centering
\renewcommand{\arraystretch}{1.5}
$\begin{array}{|c|cccc|c|}
\hline
    & \sigma_1 &\sigma_2 &\lambda_3 &\lambda_4 
& {\cal M} \left(i\frac{\sqrt{S}}{f_a}\right)\\\hline
e_r^-\,  e_s^+ \rightarrow Z \, a
& - & - & 0 & 0 &  
\frac{1}{\sqrt{2}} \left(C_{a\Phi}-C_{e\Phi}\right)\, (Y^*_e)_{rs} \\
& + & + & 0 & 0 &  
\frac{1}{\sqrt{2}} \left(C_{a\Phi}-C_{e\Phi}\right)\, (Y_e)_{sr} \\
\hline
\nu_r \, e^+_s \rightarrow W^+ \, a & - & - & 0 & 0 &  
\left(C_{a\Phi}-C_{e\Phi}\right)\, (Y^*_e)_{rs} \\\hline
u_r^a\,  \bar u_s^b \rightarrow Z \, a
& - & - & 0 & 0 &  
\frac{1}{\sqrt{2}} \left(C_{a\Phi}+C_{u\Phi}\right)\, (Y^*_u)_{rs}\,\delta_{ab}
\\
& + & + & 0 & 0 &  
\frac{1}{\sqrt{2}} \left(C_{a\Phi}+C_{u\Phi}\right)\, (Y_u)_{sr}\,\delta_{ab} \\
\hline
d_r^a\,  \bar d_s^b \rightarrow Z \, a
& - & - & 0 & 0 &  
\frac{1}{\sqrt{2}} \left(C_{a\Phi}-C_{d\Phi}\right)\, (Y^*_d)_{rs}\,\delta_{ab} \\
& + & + & 0 & 0 &  
\frac{1}{\sqrt{2}} \left(C_{a\Phi}-C_{d\Phi}\right)\, (Y_d)_{sr}\,\delta_{ab} \\
\hline
u_r^a \, \bar d_s^b \rightarrow W^+ \, a
& - & - & 0 & 0 &   \left(C_{a\Phi}-C_{d\Phi}\right)\, (Y^*_d)_{rs} \,\delta_{ab}\\
& + & + & 0 & 0 &   \left(C_{a\Phi}+C_{u\Phi}\right)\, (Y_u)_{sr} \,\delta_{ab}\\
\hline
\end{array}
  $
  \caption{ Leading contributions to the helicity amplitudes for the
    channels with fermion scattering. $a,b$ denote color indices and
    $r,s$ denote flavour indices.}
     \label{tab:fermions}
\end{table}

\clearpage
\bibliography{bibliography}

\begin{thebibliography}{93}
\expandafter\ifx\csname natexlab\endcsname\relax\def\natexlab#1{#1}\fi
\expandafter\ifx\csname bibnamefont\endcsname\relax
  \def\bibnamefont#1{#1}\fi
\expandafter\ifx\csname bibfnamefont\endcsname\relax
  \def\bibfnamefont#1{#1}\fi
\expandafter\ifx\csname citenamefont\endcsname\relax
  \def\citenamefont#1{#1}\fi
\expandafter\ifx\csname url\endcsname\relax
  \def\url#1{\texttt{#1}}\fi
\expandafter\ifx\csname urlprefix\endcsname\relax\def\urlprefix{URL }\fi
\providecommand{\bibinfo}[2]{#2}
\providecommand{\eprint}[2][]{\url{#2}}

\bibitem[{\citenamefont{Peccei and Quinn}(1977{\natexlab{a}})}]{Peccei:1977hh}
\bibinfo{author}{\bibfnamefont{R.~D.} \bibnamefont{Peccei}} \bibnamefont{and}
  \bibinfo{author}{\bibfnamefont{H.~R.} \bibnamefont{Quinn}},
  \bibinfo{journal}{Phys. Rev. Lett.} \textbf{\bibinfo{volume}{38}},
  \bibinfo{pages}{1440} (\bibinfo{year}{1977}{\natexlab{a}}).

\bibitem[{\citenamefont{Peccei and Quinn}(1977{\natexlab{b}})}]{Peccei:1977ur}
\bibinfo{author}{\bibfnamefont{R.~D.} \bibnamefont{Peccei}} \bibnamefont{and}
  \bibinfo{author}{\bibfnamefont{H.~R.} \bibnamefont{Quinn}},
  \bibinfo{journal}{Phys. Rev. D} \textbf{\bibinfo{volume}{16}},
  \bibinfo{pages}{1791} (\bibinfo{year}{1977}{\natexlab{b}}).

\bibitem[{\citenamefont{Weinberg}(1978)}]{Weinberg:1977ma}
\bibinfo{author}{\bibfnamefont{S.}~\bibnamefont{Weinberg}},
  \bibinfo{journal}{Phys. Rev. Lett.} \textbf{\bibinfo{volume}{40}},
  \bibinfo{pages}{223} (\bibinfo{year}{1978}).

\bibitem[{\citenamefont{Wilczek}(1978)}]{Wilczek:1977pj}
\bibinfo{author}{\bibfnamefont{F.}~\bibnamefont{Wilczek}},
  \bibinfo{journal}{Phys. Rev. Lett.} \textbf{\bibinfo{volume}{40}},
  \bibinfo{pages}{279} (\bibinfo{year}{1978}).

\bibitem[{\citenamefont{Kim}(1979)}]{Kim:1979if}
\bibinfo{author}{\bibfnamefont{J.~E.} \bibnamefont{Kim}},
  \bibinfo{journal}{Phys. Rev. Lett.} \textbf{\bibinfo{volume}{43}},
  \bibinfo{pages}{103} (\bibinfo{year}{1979}).

\bibitem[{\citenamefont{Shifman et~al.}(1980)\citenamefont{Shifman, Vainshtein,
  and Zakharov}}]{Shifman:1979if}
\bibinfo{author}{\bibfnamefont{M.~A.} \bibnamefont{Shifman}},
  \bibinfo{author}{\bibfnamefont{A.~I.} \bibnamefont{Vainshtein}},
  \bibnamefont{and} \bibinfo{author}{\bibfnamefont{V.~I.}
  \bibnamefont{Zakharov}}, \bibinfo{journal}{Nucl. Phys. B}
  \textbf{\bibinfo{volume}{166}}, \bibinfo{pages}{493} (\bibinfo{year}{1980}).

\bibitem[{\citenamefont{Dine et~al.}(1981)\citenamefont{Dine, Fischler, and
  Srednicki}}]{Dine:1981rt}
\bibinfo{author}{\bibfnamefont{M.}~\bibnamefont{Dine}},
  \bibinfo{author}{\bibfnamefont{W.}~\bibnamefont{Fischler}}, \bibnamefont{and}
  \bibinfo{author}{\bibfnamefont{M.}~\bibnamefont{Srednicki}},
  \bibinfo{journal}{Phys. Lett. B} \textbf{\bibinfo{volume}{104}},
  \bibinfo{pages}{199} (\bibinfo{year}{1981}).

\bibitem[{\citenamefont{Zhitnitsky}(1980)}]{Zhitnitsky:1980tq}
\bibinfo{author}{\bibfnamefont{A.~R.} \bibnamefont{Zhitnitsky}},
  \bibinfo{journal}{Sov. J. Nucl. Phys.} \textbf{\bibinfo{volume}{31}},
  \bibinfo{pages}{260} (\bibinfo{year}{1980}).

\bibitem[{\citenamefont{Rubakov}(1997)}]{Rubakov:1997vp}
\bibinfo{author}{\bibfnamefont{V.~A.} \bibnamefont{Rubakov}},
  \bibinfo{journal}{JETP Lett.} \textbf{\bibinfo{volume}{65}},
  \bibinfo{pages}{621} (\bibinfo{year}{1997}), \eprint{hep-ph/9703409}.

\bibitem[{\citenamefont{Berezhiani et~al.}(2001)\citenamefont{Berezhiani,
  Gianfagna, and Giannotti}}]{Berezhiani:2000gh}
\bibinfo{author}{\bibfnamefont{Z.}~\bibnamefont{Berezhiani}},
  \bibinfo{author}{\bibfnamefont{L.}~\bibnamefont{Gianfagna}},
  \bibnamefont{and}
  \bibinfo{author}{\bibfnamefont{M.}~\bibnamefont{Giannotti}},
  \bibinfo{journal}{Phys. Lett. B} \textbf{\bibinfo{volume}{500}},
  \bibinfo{pages}{286} (\bibinfo{year}{2001}), \eprint{hep-ph/0009290}.

\bibitem[{\citenamefont{Hsu and Sannino}(2005)}]{Hsu:2004mf}
\bibinfo{author}{\bibfnamefont{S.~D.~H.} \bibnamefont{Hsu}} \bibnamefont{and}
  \bibinfo{author}{\bibfnamefont{F.}~\bibnamefont{Sannino}},
  \bibinfo{journal}{Phys. Lett. B} \textbf{\bibinfo{volume}{605}},
  \bibinfo{pages}{369} (\bibinfo{year}{2005}), \eprint{hep-ph/0408319}.

\bibitem[{\citenamefont{Hook}(2015)}]{Hook:2014cda}
\bibinfo{author}{\bibfnamefont{A.}~\bibnamefont{Hook}}, \bibinfo{journal}{Phys.
  Rev. Lett.} \textbf{\bibinfo{volume}{114}}, \bibinfo{pages}{141801}
  (\bibinfo{year}{2015}), \eprint{1411.3325}.

\bibitem[{\citenamefont{Fukuda et~al.}(2015)\citenamefont{Fukuda, Harigaya,
  Ibe, and Yanagida}}]{Fukuda:2015ana}
\bibinfo{author}{\bibfnamefont{H.}~\bibnamefont{Fukuda}},
  \bibinfo{author}{\bibfnamefont{K.}~\bibnamefont{Harigaya}},
  \bibinfo{author}{\bibfnamefont{M.}~\bibnamefont{Ibe}}, \bibnamefont{and}
  \bibinfo{author}{\bibfnamefont{T.~T.} \bibnamefont{Yanagida}},
  \bibinfo{journal}{Phys. Rev. D} \textbf{\bibinfo{volume}{92}},
  \bibinfo{pages}{015021} (\bibinfo{year}{2015}), \eprint{1504.06084}.

\bibitem[{\citenamefont{Chiang et~al.}(2016)\citenamefont{Chiang, Fukuda, Ibe,
  and Yanagida}}]{Chiang:2016eav}
\bibinfo{author}{\bibfnamefont{C.-W.} \bibnamefont{Chiang}},
  \bibinfo{author}{\bibfnamefont{H.}~\bibnamefont{Fukuda}},
  \bibinfo{author}{\bibfnamefont{M.}~\bibnamefont{Ibe}}, \bibnamefont{and}
  \bibinfo{author}{\bibfnamefont{T.~T.} \bibnamefont{Yanagida}},
  \bibinfo{journal}{Phys. Rev. D} \textbf{\bibinfo{volume}{93}},
  \bibinfo{pages}{095016} (\bibinfo{year}{2016}), \eprint{1602.07909}.

\bibitem[{\citenamefont{Gherghetta et~al.}(2016)\citenamefont{Gherghetta,
  Nagata, and Shifman}}]{Gherghetta:2016fhp}
\bibinfo{author}{\bibfnamefont{T.}~\bibnamefont{Gherghetta}},
  \bibinfo{author}{\bibfnamefont{N.}~\bibnamefont{Nagata}}, \bibnamefont{and}
  \bibinfo{author}{\bibfnamefont{M.}~\bibnamefont{Shifman}},
  \bibinfo{journal}{Phys. Rev. D} \textbf{\bibinfo{volume}{93}},
  \bibinfo{pages}{115010} (\bibinfo{year}{2016}), \eprint{1604.01127}.

\bibitem[{\citenamefont{Dimopoulos et~al.}(2016)\citenamefont{Dimopoulos, Hook,
  Huang, and Marques-Tavares}}]{Dimopoulos:2016lvn}
\bibinfo{author}{\bibfnamefont{S.}~\bibnamefont{Dimopoulos}},
  \bibinfo{author}{\bibfnamefont{A.}~\bibnamefont{Hook}},
  \bibinfo{author}{\bibfnamefont{J.}~\bibnamefont{Huang}}, \bibnamefont{and}
  \bibinfo{author}{\bibfnamefont{G.}~\bibnamefont{Marques-Tavares}},
  \bibinfo{journal}{JHEP} \textbf{\bibinfo{volume}{11}}, \bibinfo{pages}{052}
  (\bibinfo{year}{2016}), \eprint{1606.03097}.

\bibitem[{\citenamefont{Kobakhidze}(2016)}]{Kobakhidze:2016rwh}
\bibinfo{author}{\bibfnamefont{A.}~\bibnamefont{Kobakhidze}}
  (\bibinfo{year}{2016}), \eprint{1607.06552}.

\bibitem[{\citenamefont{Agrawal and
  Howe}(2018{\natexlab{a}})}]{Agrawal:2017ksf}
\bibinfo{author}{\bibfnamefont{P.}~\bibnamefont{Agrawal}} \bibnamefont{and}
  \bibinfo{author}{\bibfnamefont{K.}~\bibnamefont{Howe}},
  \bibinfo{journal}{JHEP} \textbf{\bibinfo{volume}{12}}, \bibinfo{pages}{029}
  (\bibinfo{year}{2018}{\natexlab{a}}), \eprint{1710.04213}.

\bibitem[{\citenamefont{Agrawal and
  Howe}(2018{\natexlab{b}})}]{Agrawal:2017evu}
\bibinfo{author}{\bibfnamefont{P.}~\bibnamefont{Agrawal}} \bibnamefont{and}
  \bibinfo{author}{\bibfnamefont{K.}~\bibnamefont{Howe}},
  \bibinfo{journal}{JHEP} \textbf{\bibinfo{volume}{12}}, \bibinfo{pages}{035}
  (\bibinfo{year}{2018}{\natexlab{b}}), \eprint{1712.05803}.

\bibitem[{\citenamefont{Gaillard et~al.}(2018)\citenamefont{Gaillard, Gavela,
  Houtz, Quilez, and Del~Rey}}]{Gaillard:2018xgk}
\bibinfo{author}{\bibfnamefont{M.~K.} \bibnamefont{Gaillard}},
  \bibinfo{author}{\bibfnamefont{M.~B.} \bibnamefont{Gavela}},
  \bibinfo{author}{\bibfnamefont{R.}~\bibnamefont{Houtz}},
  \bibinfo{author}{\bibfnamefont{P.}~\bibnamefont{Quilez}}, \bibnamefont{and}
  \bibinfo{author}{\bibfnamefont{R.}~\bibnamefont{Del~Rey}},
  \bibinfo{journal}{Eur. Phys. J. C} \textbf{\bibinfo{volume}{78}},
  \bibinfo{pages}{972} (\bibinfo{year}{2018}), \eprint{1805.06465}.

\bibitem[{\citenamefont{Buen-Abad and Fan}(2019)}]{Buen-Abad:2019uoc}
\bibinfo{author}{\bibfnamefont{M.~A.} \bibnamefont{Buen-Abad}}
  \bibnamefont{and} \bibinfo{author}{\bibfnamefont{J.}~\bibnamefont{Fan}},
  \bibinfo{journal}{JHEP} \textbf{\bibinfo{volume}{12}}, \bibinfo{pages}{161}
  (\bibinfo{year}{2019}), \eprint{1911.05737}.

\bibitem[{\citenamefont{Cs\'aki et~al.}(2020)\citenamefont{Cs\'aki, Ruhdorfer,
  and Shirman}}]{Csaki:2019vte}
\bibinfo{author}{\bibfnamefont{C.}~\bibnamefont{Cs\'aki}},
  \bibinfo{author}{\bibfnamefont{M.}~\bibnamefont{Ruhdorfer}},
  \bibnamefont{and} \bibinfo{author}{\bibfnamefont{Y.}~\bibnamefont{Shirman}},
  \bibinfo{journal}{JHEP} \textbf{\bibinfo{volume}{04}}, \bibinfo{pages}{031}
  (\bibinfo{year}{2020}), \eprint{1912.02197}.

\bibitem[{\citenamefont{Hook et~al.}(2020)\citenamefont{Hook, Kumar, Liu, and
  Sundrum}}]{Hook:2019qoh}
\bibinfo{author}{\bibfnamefont{A.}~\bibnamefont{Hook}},
  \bibinfo{author}{\bibfnamefont{S.}~\bibnamefont{Kumar}},
  \bibinfo{author}{\bibfnamefont{Z.}~\bibnamefont{Liu}}, \bibnamefont{and}
  \bibinfo{author}{\bibfnamefont{R.}~\bibnamefont{Sundrum}},
  \bibinfo{journal}{Phys. Rev. Lett.} \textbf{\bibinfo{volume}{124}},
  \bibinfo{pages}{221801} (\bibinfo{year}{2020}), \eprint{1911.12364}.

\bibitem[{\citenamefont{Gherghetta and Nguyen}(2020)}]{Gherghetta:2020ofz}
\bibinfo{author}{\bibfnamefont{T.}~\bibnamefont{Gherghetta}} \bibnamefont{and}
  \bibinfo{author}{\bibfnamefont{M.~D.} \bibnamefont{Nguyen}},
  \bibinfo{journal}{JHEP} \textbf{\bibinfo{volume}{12}}, \bibinfo{pages}{094}
  (\bibinfo{year}{2020}), \eprint{2007.10875}.

\bibitem[{\citenamefont{Hook}(2018)}]{Hook:2018jle}
\bibinfo{author}{\bibfnamefont{A.}~\bibnamefont{Hook}}, \bibinfo{journal}{Phys.
  Rev. Lett.} \textbf{\bibinfo{volume}{120}}, \bibinfo{pages}{261802}
  (\bibinfo{year}{2018}), \eprint{1802.10093}.

\bibitem[{\citenamefont{Di~Luzio
  et~al.}(2021{\natexlab{a}})\citenamefont{Di~Luzio, Gavela, Quilez, and
  Ringwald}}]{DiLuzio:2021pxd}
\bibinfo{author}{\bibfnamefont{L.}~\bibnamefont{Di~Luzio}},
  \bibinfo{author}{\bibfnamefont{B.}~\bibnamefont{Gavela}},
  \bibinfo{author}{\bibfnamefont{P.}~\bibnamefont{Quilez}}, \bibnamefont{and}
  \bibinfo{author}{\bibfnamefont{A.}~\bibnamefont{Ringwald}},
  \bibinfo{journal}{JHEP} \textbf{\bibinfo{volume}{05}}, \bibinfo{pages}{184}
  (\bibinfo{year}{2021}{\natexlab{a}}), \eprint{2102.00012}.

\bibitem[{\citenamefont{Di~Luzio
  et~al.}(2021{\natexlab{b}})\citenamefont{Di~Luzio, Gavela, Quilez, and
  Ringwald}}]{DiLuzio:2021gos}
\bibinfo{author}{\bibfnamefont{L.}~\bibnamefont{Di~Luzio}},
  \bibinfo{author}{\bibfnamefont{B.}~\bibnamefont{Gavela}},
  \bibinfo{author}{\bibfnamefont{P.}~\bibnamefont{Quilez}}, \bibnamefont{and}
  \bibinfo{author}{\bibfnamefont{A.}~\bibnamefont{Ringwald}}
  (\bibinfo{year}{2021}{\natexlab{b}}), \eprint{2102.01082}.

\bibitem[{\citenamefont{Gripaios et~al.}(2009)\citenamefont{Gripaios, Pomarol,
  Riva, and Serra}}]{Gripaios:2009pe}
\bibinfo{author}{\bibfnamefont{B.}~\bibnamefont{Gripaios}},
  \bibinfo{author}{\bibfnamefont{A.}~\bibnamefont{Pomarol}},
  \bibinfo{author}{\bibfnamefont{F.}~\bibnamefont{Riva}}, \bibnamefont{and}
  \bibinfo{author}{\bibfnamefont{J.}~\bibnamefont{Serra}},
  \bibinfo{journal}{JHEP} \textbf{\bibinfo{volume}{04}}, \bibinfo{pages}{070}
  (\bibinfo{year}{2009}), \eprint{0902.1483}.

\bibitem[{\citenamefont{Gripaios et~al.}(2017)\citenamefont{Gripaios,
  Nardecchia, and You}}]{Gripaios:2016mmi}
\bibinfo{author}{\bibfnamefont{B.}~\bibnamefont{Gripaios}},
  \bibinfo{author}{\bibfnamefont{M.}~\bibnamefont{Nardecchia}},
  \bibnamefont{and} \bibinfo{author}{\bibfnamefont{T.}~\bibnamefont{You}},
  \bibinfo{journal}{Eur. Phys. J. C} \textbf{\bibinfo{volume}{77}},
  \bibinfo{pages}{28} (\bibinfo{year}{2017}), \eprint{1605.09647}.

\bibitem[{\citenamefont{Merlo et~al.}(2018)\citenamefont{Merlo, Pobbe, and
  Rigolin}}]{Merlo:2017sun}
\bibinfo{author}{\bibfnamefont{L.}~\bibnamefont{Merlo}},
  \bibinfo{author}{\bibfnamefont{F.}~\bibnamefont{Pobbe}}, \bibnamefont{and}
  \bibinfo{author}{\bibfnamefont{S.}~\bibnamefont{Rigolin}},
  \bibinfo{journal}{Eur. Phys. J. C} \textbf{\bibinfo{volume}{78}},
  \bibinfo{pages}{415} (\bibinfo{year}{2018}), \bibinfo{note}{[Erratum:
  Eur.Phys.J.C 79, 963 (2019)]}, \eprint{1710.10500}.

\bibitem[{\citenamefont{Chala et~al.}(2017)\citenamefont{Chala, Durieux,
  Grojean, de~Lima, and Matsedonskyi}}]{Chala:2017sjk}
\bibinfo{author}{\bibfnamefont{M.}~\bibnamefont{Chala}},
  \bibinfo{author}{\bibfnamefont{G.}~\bibnamefont{Durieux}},
  \bibinfo{author}{\bibfnamefont{C.}~\bibnamefont{Grojean}},
  \bibinfo{author}{\bibfnamefont{L.}~\bibnamefont{de~Lima}}, \bibnamefont{and}
  \bibinfo{author}{\bibfnamefont{O.}~\bibnamefont{Matsedonskyi}},
  \bibinfo{journal}{JHEP} \textbf{\bibinfo{volume}{06}}, \bibinfo{pages}{088}
  (\bibinfo{year}{2017}), \eprint{1703.10624}.

\bibitem[{\citenamefont{Brivio et~al.}(2019)\citenamefont{Brivio, Gavela,
  Pascoli, del Rey, and Saa}}]{Brivio:2017sdm}
\bibinfo{author}{\bibfnamefont{I.}~\bibnamefont{Brivio}},
  \bibinfo{author}{\bibfnamefont{M.~B.} \bibnamefont{Gavela}},
  \bibinfo{author}{\bibfnamefont{S.}~\bibnamefont{Pascoli}},
  \bibinfo{author}{\bibfnamefont{R.}~\bibnamefont{del Rey}}, \bibnamefont{and}
  \bibinfo{author}{\bibfnamefont{S.}~\bibnamefont{Saa}},
  \bibinfo{journal}{Chin. J. Phys.} \textbf{\bibinfo{volume}{61}},
  \bibinfo{pages}{55} (\bibinfo{year}{2019}), \eprint{1710.07715}.

\bibitem[{\citenamefont{Chikashige et~al.}(1981)\citenamefont{Chikashige,
  Mohapatra, and Peccei}}]{Chikashige:1980ui}
\bibinfo{author}{\bibfnamefont{Y.}~\bibnamefont{Chikashige}},
  \bibinfo{author}{\bibfnamefont{R.~N.} \bibnamefont{Mohapatra}},
  \bibnamefont{and} \bibinfo{author}{\bibfnamefont{R.~D.}
  \bibnamefont{Peccei}}, \bibinfo{journal}{Phys. Lett. B}
  \textbf{\bibinfo{volume}{98}}, \bibinfo{pages}{265} (\bibinfo{year}{1981}).

\bibitem[{\citenamefont{Gelmini and Roncadelli}(1981)}]{Gelmini:1980re}
\bibinfo{author}{\bibfnamefont{G.~B.} \bibnamefont{Gelmini}} \bibnamefont{and}
  \bibinfo{author}{\bibfnamefont{M.}~\bibnamefont{Roncadelli}},
  \bibinfo{journal}{Phys. Lett. B} \textbf{\bibinfo{volume}{99}},
  \bibinfo{pages}{411} (\bibinfo{year}{1981}).

\bibitem[{\citenamefont{Davidson and Wali}(1982)}]{Davidson:1981zd}
\bibinfo{author}{\bibfnamefont{A.}~\bibnamefont{Davidson}} \bibnamefont{and}
  \bibinfo{author}{\bibfnamefont{K.~C.} \bibnamefont{Wali}},
  \bibinfo{journal}{Phys. Rev. Lett.} \textbf{\bibinfo{volume}{48}},
  \bibinfo{pages}{11} (\bibinfo{year}{1982}).

\bibitem[{\citenamefont{Wilczek}(1982)}]{Wilczek:1982rv}
\bibinfo{author}{\bibfnamefont{F.}~\bibnamefont{Wilczek}},
  \bibinfo{journal}{Phys. Rev. Lett.} \textbf{\bibinfo{volume}{49}},
  \bibinfo{pages}{1549} (\bibinfo{year}{1982}).

\bibitem[{\citenamefont{Calibbi et~al.}(2017)\citenamefont{Calibbi, Goertz,
  Redigolo, Ziegler, and Zupan}}]{Calibbi:2016hwq}
\bibinfo{author}{\bibfnamefont{L.}~\bibnamefont{Calibbi}},
  \bibinfo{author}{\bibfnamefont{F.}~\bibnamefont{Goertz}},
  \bibinfo{author}{\bibfnamefont{D.}~\bibnamefont{Redigolo}},
  \bibinfo{author}{\bibfnamefont{R.}~\bibnamefont{Ziegler}}, \bibnamefont{and}
  \bibinfo{author}{\bibfnamefont{J.}~\bibnamefont{Zupan}},
  \bibinfo{journal}{Phys. Rev. D} \textbf{\bibinfo{volume}{95}},
  \bibinfo{pages}{095009} (\bibinfo{year}{2017}), \eprint{1612.08040}.

\bibitem[{\citenamefont{Ema et~al.}(2017)\citenamefont{Ema, Hamaguchi, Moroi,
  and Nakayama}}]{Ema:2016ops}
\bibinfo{author}{\bibfnamefont{Y.}~\bibnamefont{Ema}},
  \bibinfo{author}{\bibfnamefont{K.}~\bibnamefont{Hamaguchi}},
  \bibinfo{author}{\bibfnamefont{T.}~\bibnamefont{Moroi}}, \bibnamefont{and}
  \bibinfo{author}{\bibfnamefont{K.}~\bibnamefont{Nakayama}},
  \bibinfo{journal}{JHEP} \textbf{\bibinfo{volume}{01}}, \bibinfo{pages}{096}
  (\bibinfo{year}{2017}), \eprint{1612.05492}.

\bibitem[{\citenamefont{Svrcek and Witten}(2006)}]{Svrcek:2006yi}
\bibinfo{author}{\bibfnamefont{P.}~\bibnamefont{Svrcek}} \bibnamefont{and}
  \bibinfo{author}{\bibfnamefont{E.}~\bibnamefont{Witten}},
  \bibinfo{journal}{JHEP} \textbf{\bibinfo{volume}{06}}, \bibinfo{pages}{051}
  (\bibinfo{year}{2006}), \eprint{hep-th/0605206}.

\bibitem[{\citenamefont{Arvanitaki et~al.}(2010)\citenamefont{Arvanitaki,
  Dimopoulos, Dubovsky, Kaloper, and March-Russell}}]{Arvanitaki:2009fg}
\bibinfo{author}{\bibfnamefont{A.}~\bibnamefont{Arvanitaki}},
  \bibinfo{author}{\bibfnamefont{S.}~\bibnamefont{Dimopoulos}},
  \bibinfo{author}{\bibfnamefont{S.}~\bibnamefont{Dubovsky}},
  \bibinfo{author}{\bibfnamefont{N.}~\bibnamefont{Kaloper}}, \bibnamefont{and}
  \bibinfo{author}{\bibfnamefont{J.}~\bibnamefont{March-Russell}},
  \bibinfo{journal}{Phys. Rev. D} \textbf{\bibinfo{volume}{81}},
  \bibinfo{pages}{123530} (\bibinfo{year}{2010}), \eprint{0905.4720}.

\bibitem[{\citenamefont{Georgi et~al.}(1986)\citenamefont{Georgi, Kaplan, and
  Randall}}]{Georgi:1986df}
\bibinfo{author}{\bibfnamefont{H.}~\bibnamefont{Georgi}},
  \bibinfo{author}{\bibfnamefont{D.~B.} \bibnamefont{Kaplan}},
  \bibnamefont{and} \bibinfo{author}{\bibfnamefont{L.}~\bibnamefont{Randall}},
  \bibinfo{journal}{Phys. Lett. B} \textbf{\bibinfo{volume}{169}},
  \bibinfo{pages}{73} (\bibinfo{year}{1986}).

\bibitem[{\citenamefont{Choi et~al.}(1986)\citenamefont{Choi, Kang, and
  Kim}}]{Choi:1986zw}
\bibinfo{author}{\bibfnamefont{K.}~\bibnamefont{Choi}},
  \bibinfo{author}{\bibfnamefont{K.}~\bibnamefont{Kang}}, \bibnamefont{and}
  \bibinfo{author}{\bibfnamefont{J.~E.} \bibnamefont{Kim}},
  \bibinfo{journal}{Phys. Lett. B} \textbf{\bibinfo{volume}{181}},
  \bibinfo{pages}{145} (\bibinfo{year}{1986}).

\bibitem[{\citenamefont{Choi et~al.}(2017)\citenamefont{Choi, Im, Park, and
  Yun}}]{Choi:2017gpf}
\bibinfo{author}{\bibfnamefont{K.}~\bibnamefont{Choi}},
  \bibinfo{author}{\bibfnamefont{S.~H.} \bibnamefont{Im}},
  \bibinfo{author}{\bibfnamefont{C.~B.} \bibnamefont{Park}}, \bibnamefont{and}
  \bibinfo{author}{\bibfnamefont{S.}~\bibnamefont{Yun}},
  \bibinfo{journal}{JHEP} \textbf{\bibinfo{volume}{11}}, \bibinfo{pages}{070}
  (\bibinfo{year}{2017}), \eprint{1708.00021}.

\bibitem[{\citenamefont{Chala et~al.}(2021)\citenamefont{Chala, Guedes, Ramos,
  and Santiago}}]{Chala:2020wvs}
\bibinfo{author}{\bibfnamefont{M.}~\bibnamefont{Chala}},
  \bibinfo{author}{\bibfnamefont{G.}~\bibnamefont{Guedes}},
  \bibinfo{author}{\bibfnamefont{M.}~\bibnamefont{Ramos}}, \bibnamefont{and}
  \bibinfo{author}{\bibfnamefont{J.}~\bibnamefont{Santiago}},
  \bibinfo{journal}{Eur. Phys. J. C} \textbf{\bibinfo{volume}{81}},
  \bibinfo{pages}{181} (\bibinfo{year}{2021}), \eprint{2012.09017}.

\bibitem[{\citenamefont{Bauer et~al.}(2021)\citenamefont{Bauer, Neubert,
  Renner, Schnubel, and Thamm}}]{Bauer:2020jbp}
\bibinfo{author}{\bibfnamefont{M.}~\bibnamefont{Bauer}},
  \bibinfo{author}{\bibfnamefont{M.}~\bibnamefont{Neubert}},
  \bibinfo{author}{\bibfnamefont{S.}~\bibnamefont{Renner}},
  \bibinfo{author}{\bibfnamefont{M.}~\bibnamefont{Schnubel}}, \bibnamefont{and}
  \bibinfo{author}{\bibfnamefont{A.}~\bibnamefont{Thamm}},
  \bibinfo{journal}{JHEP} \textbf{\bibinfo{volume}{04}}, \bibinfo{pages}{063}
  (\bibinfo{year}{2021}), \eprint{2012.12272}.

\bibitem[{\citenamefont{Galda et~al.}(2021)\citenamefont{Galda, Neubert, and
  Renner}}]{Galda:2021hbr}
\bibinfo{author}{\bibfnamefont{A.~M.} \bibnamefont{Galda}},
  \bibinfo{author}{\bibfnamefont{M.}~\bibnamefont{Neubert}}, \bibnamefont{and}
  \bibinfo{author}{\bibfnamefont{S.}~\bibnamefont{Renner}}
  (\bibinfo{year}{2021}), \eprint{2105.01078}.

\bibitem[{\citenamefont{Alonso-\'Alvarez
  et~al.}(2021)\citenamefont{Alonso-\'Alvarez, Ertas, Jaeckel, Kahlhoefer, and
  Thormaehlen}}]{Alonso-Alvarez:2021ett}
\bibinfo{author}{\bibfnamefont{G.}~\bibnamefont{Alonso-\'Alvarez}},
  \bibinfo{author}{\bibfnamefont{F.}~\bibnamefont{Ertas}},
  \bibinfo{author}{\bibfnamefont{J.}~\bibnamefont{Jaeckel}},
  \bibinfo{author}{\bibfnamefont{F.}~\bibnamefont{Kahlhoefer}},
  \bibnamefont{and} \bibinfo{author}{\bibfnamefont{L.~J.}
  \bibnamefont{Thormaehlen}} (\bibinfo{year}{2021}), \eprint{2101.03173}.

\bibitem[{\citenamefont{Lee et~al.}(1977{\natexlab{a}})\citenamefont{Lee,
  Quigg, and Thacker}}]{Lee:1977yc}
\bibinfo{author}{\bibfnamefont{B.~W.} \bibnamefont{Lee}},
  \bibinfo{author}{\bibfnamefont{C.}~\bibnamefont{Quigg}}, \bibnamefont{and}
  \bibinfo{author}{\bibfnamefont{H.~B.} \bibnamefont{Thacker}},
  \bibinfo{journal}{Phys. Rev. Lett.} \textbf{\bibinfo{volume}{38}},
  \bibinfo{pages}{883} (\bibinfo{year}{1977}{\natexlab{a}}).

\bibitem[{\citenamefont{Lee et~al.}(1977{\natexlab{b}})\citenamefont{Lee,
  Quigg, and Thacker}}]{Lee:1977eg}
\bibinfo{author}{\bibfnamefont{B.~W.} \bibnamefont{Lee}},
  \bibinfo{author}{\bibfnamefont{C.}~\bibnamefont{Quigg}}, \bibnamefont{and}
  \bibinfo{author}{\bibfnamefont{H.~B.} \bibnamefont{Thacker}},
  \bibinfo{journal}{Phys. Rev. D} \textbf{\bibinfo{volume}{16}},
  \bibinfo{pages}{1519} (\bibinfo{year}{1977}{\natexlab{b}}).

\bibitem[{\citenamefont{Chanowitz et~al.}(1978)\citenamefont{Chanowitz, Furman,
  and Hinchliffe}}]{Chanowitz:1978uj}
\bibinfo{author}{\bibfnamefont{M.~S.} \bibnamefont{Chanowitz}},
  \bibinfo{author}{\bibfnamefont{M.~A.} \bibnamefont{Furman}},
  \bibnamefont{and}
  \bibinfo{author}{\bibfnamefont{I.}~\bibnamefont{Hinchliffe}},
  \bibinfo{journal}{Phys. Lett. B} \textbf{\bibinfo{volume}{78}},
  \bibinfo{pages}{285} (\bibinfo{year}{1978}).

\bibitem[{\citenamefont{Cornwall et~al.}(1974)\citenamefont{Cornwall, Levin,
  and Tiktopoulos}}]{Cornwall:1974km}
\bibinfo{author}{\bibfnamefont{J.~M.} \bibnamefont{Cornwall}},
  \bibinfo{author}{\bibfnamefont{D.~N.} \bibnamefont{Levin}}, \bibnamefont{and}
  \bibinfo{author}{\bibfnamefont{G.}~\bibnamefont{Tiktopoulos}},
  \bibinfo{journal}{Phys. Rev. D} \textbf{\bibinfo{volume}{10}},
  \bibinfo{pages}{1145} (\bibinfo{year}{1974}), \bibinfo{note}{[Erratum:
  Phys.Rev.D 11, 972 (1975)]}.

\bibitem[{\citenamefont{Bilchak et~al.}(1988)\citenamefont{Bilchak, Kuroda, and
  Schildknecht}}]{Bilchak:1987cp}
\bibinfo{author}{\bibfnamefont{C.}~\bibnamefont{Bilchak}},
  \bibinfo{author}{\bibfnamefont{M.}~\bibnamefont{Kuroda}}, \bibnamefont{and}
  \bibinfo{author}{\bibfnamefont{D.}~\bibnamefont{Schildknecht}},
  \bibinfo{journal}{Nucl. Phys.} \textbf{\bibinfo{volume}{B299}},
  \bibinfo{pages}{7} (\bibinfo{year}{1988}).

\bibitem[{\citenamefont{Gounaris et~al.}(1994)\citenamefont{Gounaris, Layssac,
  and Renard}}]{Gounaris:1993fh}
\bibinfo{author}{\bibfnamefont{G.~J.} \bibnamefont{Gounaris}},
  \bibinfo{author}{\bibfnamefont{J.}~\bibnamefont{Layssac}}, \bibnamefont{and}
  \bibinfo{author}{\bibfnamefont{F.~M.} \bibnamefont{Renard}},
  \bibinfo{journal}{Phys. Lett.} \textbf{\bibinfo{volume}{B332}},
  \bibinfo{pages}{146} (\bibinfo{year}{1994}), \eprint{hep-ph/9311370}.

\bibitem[{\citenamefont{Gounaris
  et~al.}(1995{\natexlab{a}})\citenamefont{Gounaris, Layssac, Paschalis, and
  Renard}}]{Gounaris:1994cm}
\bibinfo{author}{\bibfnamefont{G.~J.} \bibnamefont{Gounaris}},
  \bibinfo{author}{\bibfnamefont{J.}~\bibnamefont{Layssac}},
  \bibinfo{author}{\bibfnamefont{J.~E.} \bibnamefont{Paschalis}},
  \bibnamefont{and} \bibinfo{author}{\bibfnamefont{F.~M.}
  \bibnamefont{Renard}}, \bibinfo{journal}{Z. Phys.}
  \textbf{\bibinfo{volume}{C66}}, \bibinfo{pages}{619}
  (\bibinfo{year}{1995}{\natexlab{a}}), \eprint{hep-ph/9409260}.

\bibitem[{\citenamefont{Gounaris
  et~al.}(1995{\natexlab{b}})\citenamefont{Gounaris, Renard, and
  Tsirigoti}}]{Gounaris:1995ed}
\bibinfo{author}{\bibfnamefont{G.~J.} \bibnamefont{Gounaris}},
  \bibinfo{author}{\bibfnamefont{F.~M.} \bibnamefont{Renard}},
  \bibnamefont{and}
  \bibinfo{author}{\bibfnamefont{G.}~\bibnamefont{Tsirigoti}},
  \bibinfo{journal}{Phys. Lett.} \textbf{\bibinfo{volume}{B350}},
  \bibinfo{pages}{212} (\bibinfo{year}{1995}{\natexlab{b}}),
  \eprint{hep-ph/9502376}.

\bibitem[{\citenamefont{Degrande}(2013)}]{Degrande:2013mh}
\bibinfo{author}{\bibfnamefont{C.}~\bibnamefont{Degrande}},
  \bibinfo{journal}{EPJ Web Conf.} \textbf{\bibinfo{volume}{49}},
  \bibinfo{pages}{14009} (\bibinfo{year}{2013}), \eprint{1302.1112}.

\bibitem[{\citenamefont{Baur and Zeppenfeld}(1988)}]{Baur:1987mt}
\bibinfo{author}{\bibfnamefont{U.}~\bibnamefont{Baur}} \bibnamefont{and}
  \bibinfo{author}{\bibfnamefont{D.}~\bibnamefont{Zeppenfeld}},
  \bibinfo{journal}{Phys. Lett. B} \textbf{\bibinfo{volume}{201}},
  \bibinfo{pages}{383} (\bibinfo{year}{1988}).

\bibitem[{\citenamefont{Dahiya et~al.}(2016)\citenamefont{Dahiya, Dutta, and
  Islam}}]{Dahiya:2013uba}
\bibinfo{author}{\bibfnamefont{M.}~\bibnamefont{Dahiya}},
  \bibinfo{author}{\bibfnamefont{S.}~\bibnamefont{Dutta}}, \bibnamefont{and}
  \bibinfo{author}{\bibfnamefont{R.}~\bibnamefont{Islam}},
  \bibinfo{journal}{Phys. Rev.} \textbf{\bibinfo{volume}{D93}},
  \bibinfo{pages}{055013} (\bibinfo{year}{2016}), \eprint{1311.4523}.

\bibitem[{\citenamefont{Ghosh et~al.}(2017)\citenamefont{Ghosh, Islam, and
  Kundu}}]{Ghosh:2017coz}
\bibinfo{author}{\bibfnamefont{S.}~\bibnamefont{Ghosh}},
  \bibinfo{author}{\bibfnamefont{R.}~\bibnamefont{Islam}}, \bibnamefont{and}
  \bibinfo{author}{\bibfnamefont{A.}~\bibnamefont{Kundu}}
  (\bibinfo{year}{2017}), \eprint{1704.01867}.

\bibitem[{\citenamefont{Corbett et~al.}(2015)\citenamefont{Corbett, \'Eboli,
  and Gonzalez-Garcia}}]{Corbett:2014ora}
\bibinfo{author}{\bibfnamefont{T.}~\bibnamefont{Corbett}},
  \bibinfo{author}{\bibfnamefont{O.~J.~P.} \bibnamefont{\'Eboli}},
  \bibnamefont{and} \bibinfo{author}{\bibfnamefont{M.~C.}
  \bibnamefont{Gonzalez-Garcia}}, \bibinfo{journal}{Phys. Rev. D}
  \textbf{\bibinfo{volume}{91}}, \bibinfo{pages}{035014}
  (\bibinfo{year}{2015}), \eprint{1411.5026}.

\bibitem[{\citenamefont{Corbett et~al.}(2017)\citenamefont{Corbett, \'Eboli,
  and Gonzalez-Garcia}}]{Corbett:2017qgl}
\bibinfo{author}{\bibfnamefont{T.}~\bibnamefont{Corbett}},
  \bibinfo{author}{\bibfnamefont{O.~J.~P.} \bibnamefont{\'Eboli}},
  \bibnamefont{and} \bibinfo{author}{\bibfnamefont{M.~C.}
  \bibnamefont{Gonzalez-Garcia}}, \bibinfo{journal}{Phys. Rev. D}
  \textbf{\bibinfo{volume}{96}}, \bibinfo{pages}{035006}
  (\bibinfo{year}{2017}), \eprint{1705.09294}.

\bibitem[{\citenamefont{Almeida et~al.}(2020)\citenamefont{Almeida, \'Eboli,
  and Gonzalez\textendash{}Garcia}}]{Almeida:2020ylr}
\bibinfo{author}{\bibfnamefont{E.~d.~S.} \bibnamefont{Almeida}},
  \bibinfo{author}{\bibfnamefont{O.~J.~P.} \bibnamefont{\'Eboli}},
  \bibnamefont{and} \bibinfo{author}{\bibfnamefont{M.~C.}
  \bibnamefont{Gonzalez\textendash{}Garcia}}, \bibinfo{journal}{Phys. Rev. D}
  \textbf{\bibinfo{volume}{101}}, \bibinfo{pages}{113003}
  (\bibinfo{year}{2020}), \eprint{2004.05174}.

\bibitem[{\citenamefont{Marciano et~al.}(2016)\citenamefont{Marciano, Masiero,
  Paradisi, and Passera}}]{Marciano:2016yhf}
\bibinfo{author}{\bibfnamefont{W.~J.} \bibnamefont{Marciano}},
  \bibinfo{author}{\bibfnamefont{A.}~\bibnamefont{Masiero}},
  \bibinfo{author}{\bibfnamefont{P.}~\bibnamefont{Paradisi}}, \bibnamefont{and}
  \bibinfo{author}{\bibfnamefont{M.}~\bibnamefont{Passera}},
  \bibinfo{journal}{Phys. Rev. D} \textbf{\bibinfo{volume}{94}},
  \bibinfo{pages}{115033} (\bibinfo{year}{2016}), \eprint{1607.01022}.

\bibitem[{\citenamefont{Cornella et~al.}(2020)\citenamefont{Cornella, Paradisi,
  and Sumensari}}]{Cornella:2019uxs}
\bibinfo{author}{\bibfnamefont{C.}~\bibnamefont{Cornella}},
  \bibinfo{author}{\bibfnamefont{P.}~\bibnamefont{Paradisi}}, \bibnamefont{and}
  \bibinfo{author}{\bibfnamefont{O.}~\bibnamefont{Sumensari}},
  \bibinfo{journal}{JHEP} \textbf{\bibinfo{volume}{01}}, \bibinfo{pages}{158}
  (\bibinfo{year}{2020}), \eprint{1911.06279}.

\bibitem[{\citenamefont{Brivio et~al.}(2017)\citenamefont{Brivio, Gavela,
  Merlo, Mimasu, No, del Rey, and Sanz}}]{Brivio:2017ije}
\bibinfo{author}{\bibfnamefont{I.}~\bibnamefont{Brivio}},
  \bibinfo{author}{\bibfnamefont{M.~B.} \bibnamefont{Gavela}},
  \bibinfo{author}{\bibfnamefont{L.}~\bibnamefont{Merlo}},
  \bibinfo{author}{\bibfnamefont{K.}~\bibnamefont{Mimasu}},
  \bibinfo{author}{\bibfnamefont{J.~M.} \bibnamefont{No}},
  \bibinfo{author}{\bibfnamefont{R.}~\bibnamefont{del Rey}}, \bibnamefont{and}
  \bibinfo{author}{\bibfnamefont{V.}~\bibnamefont{Sanz}},
  \bibinfo{journal}{Eur. Phys. J. C} \textbf{\bibinfo{volume}{77}},
  \bibinfo{pages}{572} (\bibinfo{year}{2017}), \eprint{1701.05379}.

\bibitem[{\citenamefont{Bardeen et~al.}(1978)\citenamefont{Bardeen, Tye, and
  Vermaseren}}]{Bardeen:1978nq}
\bibinfo{author}{\bibfnamefont{W.~A.} \bibnamefont{Bardeen}},
  \bibinfo{author}{\bibfnamefont{S.~H.~H.} \bibnamefont{Tye}},
  \bibnamefont{and} \bibinfo{author}{\bibfnamefont{J.~A.~M.}
  \bibnamefont{Vermaseren}}, \bibinfo{journal}{Phys. Lett. B}
  \textbf{\bibinfo{volume}{76}}, \bibinfo{pages}{580} (\bibinfo{year}{1978}).

\bibitem[{\citenamefont{Di~Vecchia and Veneziano}(1980)}]{DiVecchia:1980yfw}
\bibinfo{author}{\bibfnamefont{P.}~\bibnamefont{Di~Vecchia}} \bibnamefont{and}
  \bibinfo{author}{\bibfnamefont{G.}~\bibnamefont{Veneziano}},
  \bibinfo{journal}{Nucl. Phys. B} \textbf{\bibinfo{volume}{171}},
  \bibinfo{pages}{253} (\bibinfo{year}{1980}).

\bibitem[{\citenamefont{Grilli~di Cortona et~al.}(2016)\citenamefont{Grilli~di
  Cortona, Hardy, Pardo~Vega, and Villadoro}}]{diCortona:2015ldu}
\bibinfo{author}{\bibfnamefont{G.}~\bibnamefont{Grilli~di Cortona}},
  \bibinfo{author}{\bibfnamefont{E.}~\bibnamefont{Hardy}},
  \bibinfo{author}{\bibfnamefont{J.}~\bibnamefont{Pardo~Vega}},
  \bibnamefont{and}
  \bibinfo{author}{\bibfnamefont{G.}~\bibnamefont{Villadoro}},
  \bibinfo{journal}{JHEP} \textbf{\bibinfo{volume}{01}}, \bibinfo{pages}{034}
  (\bibinfo{year}{2016}), \eprint{1511.02867}.

\bibitem[{\citenamefont{Chivukula and Georgi}(1987)}]{Chivukula:1987py}
\bibinfo{author}{\bibfnamefont{R.~S.} \bibnamefont{Chivukula}}
  \bibnamefont{and} \bibinfo{author}{\bibfnamefont{H.}~\bibnamefont{Georgi}},
  \bibinfo{journal}{Phys. Lett.} \textbf{\bibinfo{volume}{B188}},
  \bibinfo{pages}{99} (\bibinfo{year}{1987}).

\bibitem[{\citenamefont{Hall and Randall}(1990)}]{Hall:1990ac}
\bibinfo{author}{\bibfnamefont{L.~J.} \bibnamefont{Hall}} \bibnamefont{and}
  \bibinfo{author}{\bibfnamefont{L.}~\bibnamefont{Randall}},
  \bibinfo{journal}{Phys. Rev. Lett.} \textbf{\bibinfo{volume}{65}},
  \bibinfo{pages}{2939} (\bibinfo{year}{1990}).

\bibitem[{\citenamefont{D'Ambrosio et~al.}(2002)\citenamefont{D'Ambrosio,
  Giudice, Isidori, and Strumia}}]{D'Ambrosio:2002ex}
\bibinfo{author}{\bibfnamefont{G.}~\bibnamefont{D'Ambrosio}},
  \bibinfo{author}{\bibfnamefont{G.~F.} \bibnamefont{Giudice}},
  \bibinfo{author}{\bibfnamefont{G.}~\bibnamefont{Isidori}}, \bibnamefont{and}
  \bibinfo{author}{\bibfnamefont{A.}~\bibnamefont{Strumia}},
  \bibinfo{journal}{Nucl. Phys.} \textbf{\bibinfo{volume}{B645}},
  \bibinfo{pages}{155} (\bibinfo{year}{2002}), \eprint{hep-ph/0207036}.

\bibitem[{\citenamefont{Bonilla et~al.}(2021)\citenamefont{Bonilla, Brivio,
  Gavela, and Sanz}}]{Bonilla:2021ufe}
\bibinfo{author}{\bibfnamefont{J.}~\bibnamefont{Bonilla}},
  \bibinfo{author}{\bibfnamefont{I.}~\bibnamefont{Brivio}},
  \bibinfo{author}{\bibfnamefont{M.~B.} \bibnamefont{Gavela}},
  \bibnamefont{and} \bibinfo{author}{\bibfnamefont{V.}~\bibnamefont{Sanz}}
  (\bibinfo{year}{2021}), \eprint{2107.11392}.

\bibitem[{\citenamefont{Draper and McKeen}(2012)}]{Draper:2012xt}
\bibinfo{author}{\bibfnamefont{P.}~\bibnamefont{Draper}} \bibnamefont{and}
  \bibinfo{author}{\bibfnamefont{D.}~\bibnamefont{McKeen}},
  \bibinfo{journal}{Phys. Rev. D} \textbf{\bibinfo{volume}{85}},
  \bibinfo{pages}{115023} (\bibinfo{year}{2012}), \eprint{1204.1061}.

\bibitem[{\citenamefont{Bauer et~al.}(2017)\citenamefont{Bauer, Neubert, and
  Thamm}}]{Bauer:2017ris}
\bibinfo{author}{\bibfnamefont{M.}~\bibnamefont{Bauer}},
  \bibinfo{author}{\bibfnamefont{M.}~\bibnamefont{Neubert}}, \bibnamefont{and}
  \bibinfo{author}{\bibfnamefont{A.}~\bibnamefont{Thamm}},
  \bibinfo{journal}{JHEP} \textbf{\bibinfo{volume}{12}}, \bibinfo{pages}{044}
  (\bibinfo{year}{2017}), \eprint{1708.00443}.

\bibitem[{\citenamefont{Bauer et~al.}(2019)\citenamefont{Bauer, Heiles,
  Neubert, and Thamm}}]{Bauer:2018uxu}
\bibinfo{author}{\bibfnamefont{M.}~\bibnamefont{Bauer}},
  \bibinfo{author}{\bibfnamefont{M.}~\bibnamefont{Heiles}},
  \bibinfo{author}{\bibfnamefont{M.}~\bibnamefont{Neubert}}, \bibnamefont{and}
  \bibinfo{author}{\bibfnamefont{A.}~\bibnamefont{Thamm}},
  \bibinfo{journal}{Eur. Phys. J. C} \textbf{\bibinfo{volume}{79}},
  \bibinfo{pages}{74} (\bibinfo{year}{2019}), \eprint{1808.10323}.

\bibitem[{\citenamefont{Davoudiasl et~al.}(2021)\citenamefont{Davoudiasl,
  Marcarelli, Miesch, and Neil}}]{Davoudiasl:2021haa}
\bibinfo{author}{\bibfnamefont{H.}~\bibnamefont{Davoudiasl}},
  \bibinfo{author}{\bibfnamefont{R.}~\bibnamefont{Marcarelli}},
  \bibinfo{author}{\bibfnamefont{N.}~\bibnamefont{Miesch}}, \bibnamefont{and}
  \bibinfo{author}{\bibfnamefont{E.~T.} \bibnamefont{Neil}}
  (\bibinfo{year}{2021}), \eprint{2105.05866}.

\bibitem[{\citenamefont{Criado}(2019)}]{Criado:2019ugp}
\bibinfo{author}{\bibfnamefont{J.~C.} \bibnamefont{Criado}},
  \bibinfo{journal}{Eur. Phys. J. C} \textbf{\bibinfo{volume}{79}},
  \bibinfo{pages}{256} (\bibinfo{year}{2019}), \eprint{1901.03501}.

\bibitem[{\citenamefont{Jacob and Wick}(1959)}]{Jacob:1959at}
\bibinfo{author}{\bibfnamefont{M.}~\bibnamefont{Jacob}} \bibnamefont{and}
  \bibinfo{author}{\bibfnamefont{G.~C.} \bibnamefont{Wick}},
  \bibinfo{journal}{Annals Phys.} \textbf{\bibinfo{volume}{7}},
  \bibinfo{pages}{404} (\bibinfo{year}{1959}).

\bibitem[{\citenamefont{Jaeckel et~al.}(2013)\citenamefont{Jaeckel, Jankowiak,
  and Spannowsky}}]{Jaeckel:2012yz}
\bibinfo{author}{\bibfnamefont{J.}~\bibnamefont{Jaeckel}},
  \bibinfo{author}{\bibfnamefont{M.}~\bibnamefont{Jankowiak}},
  \bibnamefont{and}
  \bibinfo{author}{\bibfnamefont{M.}~\bibnamefont{Spannowsky}},
  \bibinfo{journal}{Phys. Dark Univ.} \textbf{\bibinfo{volume}{2}},
  \bibinfo{pages}{111} (\bibinfo{year}{2013}), \eprint{1212.3620}.

\bibitem[{\citenamefont{Mimasu and Sanz}(2015)}]{Mimasu:2014nea}
\bibinfo{author}{\bibfnamefont{K.}~\bibnamefont{Mimasu}} \bibnamefont{and}
  \bibinfo{author}{\bibfnamefont{V.}~\bibnamefont{Sanz}},
  \bibinfo{journal}{JHEP} \textbf{\bibinfo{volume}{06}}, \bibinfo{pages}{173}
  (\bibinfo{year}{2015}), \eprint{1409.4792}.

\bibitem[{\citenamefont{Jaeckel and Spannowsky}(2016)}]{Jaeckel:2015jla}
\bibinfo{author}{\bibfnamefont{J.}~\bibnamefont{Jaeckel}} \bibnamefont{and}
  \bibinfo{author}{\bibfnamefont{M.}~\bibnamefont{Spannowsky}},
  \bibinfo{journal}{Phys. Lett. B} \textbf{\bibinfo{volume}{753}},
  \bibinfo{pages}{482} (\bibinfo{year}{2016}), \eprint{1509.00476}.

\bibitem[{\citenamefont{Frugiuele et~al.}(2018)\citenamefont{Frugiuele, Fuchs,
  Perez, and Schlaffer}}]{Frugiuele:2018coc}
\bibinfo{author}{\bibfnamefont{C.}~\bibnamefont{Frugiuele}},
  \bibinfo{author}{\bibfnamefont{E.}~\bibnamefont{Fuchs}},
  \bibinfo{author}{\bibfnamefont{G.}~\bibnamefont{Perez}}, \bibnamefont{and}
  \bibinfo{author}{\bibfnamefont{M.}~\bibnamefont{Schlaffer}},
  \bibinfo{journal}{JHEP} \textbf{\bibinfo{volume}{10}}, \bibinfo{pages}{151}
  (\bibinfo{year}{2018}), \eprint{1807.10842}.

\bibitem[{\citenamefont{Craig et~al.}(2018)\citenamefont{Craig, Hook, and
  Kasko}}]{Craig:2018kne}
\bibinfo{author}{\bibfnamefont{N.}~\bibnamefont{Craig}},
  \bibinfo{author}{\bibfnamefont{A.}~\bibnamefont{Hook}}, \bibnamefont{and}
  \bibinfo{author}{\bibfnamefont{S.}~\bibnamefont{Kasko}},
  \bibinfo{journal}{JHEP} \textbf{\bibinfo{volume}{09}}, \bibinfo{pages}{028}
  (\bibinfo{year}{2018}), \eprint{1805.06538}.

\bibitem[{\citenamefont{Ebadi et~al.}(2019)\citenamefont{Ebadi, Khatibi, and
  Mohammadi~Najafabadi}}]{Ebadi:2019gij}
\bibinfo{author}{\bibfnamefont{J.}~\bibnamefont{Ebadi}},
  \bibinfo{author}{\bibfnamefont{S.}~\bibnamefont{Khatibi}}, \bibnamefont{and}
  \bibinfo{author}{\bibfnamefont{M.}~\bibnamefont{Mohammadi~Najafabadi}},
  \bibinfo{journal}{Phys. Rev. D} \textbf{\bibinfo{volume}{100}},
  \bibinfo{pages}{015016} (\bibinfo{year}{2019}), \eprint{1901.03061}.

\bibitem[{\citenamefont{Yue et~al.}(2019)\citenamefont{Yue, Liu, and
  Guo}}]{Yue:2019gbh}
\bibinfo{author}{\bibfnamefont{C.-X.} \bibnamefont{Yue}},
  \bibinfo{author}{\bibfnamefont{M.-Z.} \bibnamefont{Liu}}, \bibnamefont{and}
  \bibinfo{author}{\bibfnamefont{Y.-C.} \bibnamefont{Guo}},
  \bibinfo{journal}{Phys. Rev. D} \textbf{\bibinfo{volume}{100}},
  \bibinfo{pages}{015020} (\bibinfo{year}{2019}), \eprint{1904.10657}.

\bibitem[{\citenamefont{Gavela et~al.}(2020)\citenamefont{Gavela, No, Sanz, and
  de~Troc\'oniz}}]{Gavela:2019cmq}
\bibinfo{author}{\bibfnamefont{M.~B.} \bibnamefont{Gavela}},
  \bibinfo{author}{\bibfnamefont{J.~M.} \bibnamefont{No}},
  \bibinfo{author}{\bibfnamefont{V.}~\bibnamefont{Sanz}}, \bibnamefont{and}
  \bibinfo{author}{\bibfnamefont{J.~F.} \bibnamefont{de~Troc\'oniz}},
  \bibinfo{journal}{Phys. Rev. Lett.} \textbf{\bibinfo{volume}{124}},
  \bibinfo{pages}{051802} (\bibinfo{year}{2020}), \eprint{1905.12953}.

\bibitem[{\citenamefont{\.Inan and Kisselev}(2020)}]{Inan:2020aal}
\bibinfo{author}{\bibfnamefont{S.~C.} \bibnamefont{\.Inan}} \bibnamefont{and}
  \bibinfo{author}{\bibfnamefont{A.~V.} \bibnamefont{Kisselev}},
  \bibinfo{journal}{JHEP} \textbf{\bibinfo{volume}{06}}, \bibinfo{pages}{183}
  (\bibinfo{year}{2020}), \eprint{2003.01978}.

\bibitem[{\citenamefont{Haghighat et~al.}(2020)\citenamefont{Haghighat,
  Haji~Raissi, and Mohammadi~Najafabadi}}]{Haghighat:2020nuh}
\bibinfo{author}{\bibfnamefont{G.}~\bibnamefont{Haghighat}},
  \bibinfo{author}{\bibfnamefont{D.}~\bibnamefont{Haji~Raissi}},
  \bibnamefont{and}
  \bibinfo{author}{\bibfnamefont{M.}~\bibnamefont{Mohammadi~Najafabadi}},
  \bibinfo{journal}{Phys. Rev. D} \textbf{\bibinfo{volume}{102}},
  \bibinfo{pages}{115010} (\bibinfo{year}{2020}), \eprint{2006.05302}.

\bibitem[{\citenamefont{Goncalves and Sauter}(2020)}]{Goncalves:2020bqi}
\bibinfo{author}{\bibfnamefont{V.~P.} \bibnamefont{Goncalves}}
  \bibnamefont{and} \bibinfo{author}{\bibfnamefont{W.~K.}
  \bibnamefont{Sauter}}, \bibinfo{journal}{Phys. Lett. B}
  \textbf{\bibinfo{volume}{811}}, \bibinfo{pages}{135981}
  (\bibinfo{year}{2020}), \eprint{2006.16716}.

\bibitem[{\citenamefont{Fl\'orez et~al.}(2021)\citenamefont{Fl\'orez, Gurrola,
  Johns, Sheldon, Sheridan, Sinha, and Soubasis}}]{Florez:2021zoo}
\bibinfo{author}{\bibfnamefont{A.}~\bibnamefont{Fl\'orez}},
  \bibinfo{author}{\bibfnamefont{A.}~\bibnamefont{Gurrola}},
  \bibinfo{author}{\bibfnamefont{W.}~\bibnamefont{Johns}},
  \bibinfo{author}{\bibfnamefont{P.}~\bibnamefont{Sheldon}},
  \bibinfo{author}{\bibfnamefont{E.}~\bibnamefont{Sheridan}},
  \bibinfo{author}{\bibfnamefont{K.}~\bibnamefont{Sinha}}, \bibnamefont{and}
  \bibinfo{author}{\bibfnamefont{B.}~\bibnamefont{Soubasis}},
  \bibinfo{journal}{Phys. Rev. D} \textbf{\bibinfo{volume}{103}},
  \bibinfo{pages}{095001} (\bibinfo{year}{2021}), \eprint{2101.11119}.

\bibitem[{\citenamefont{Bonilla et~al.}()\citenamefont{Bonilla, Brivio,
  Machado, and F.~de Troc\'oniz}}]{VBS}
\bibinfo{author}{\bibfnamefont{J.}~\bibnamefont{Bonilla}},
  \bibinfo{author}{\bibfnamefont{I.}~\bibnamefont{Brivio}},
  \bibinfo{author}{\bibfnamefont{J.}~\bibnamefont{Machado}}, \bibnamefont{and}
  \bibinfo{author}{\bibfnamefont{J.}~\bibnamefont{F.~de Troc\'oniz}},
  \bibinfo{note}{in preparation}.

\bibitem[{\citenamefont{Aad et~al.}(2021{\natexlab{a}})}]{Aad:2021egl}
\bibinfo{author}{\bibfnamefont{G.}~\bibnamefont{Aad}} \bibnamefont{et~al.}
  (\bibinfo{collaboration}{ATLAS}) (\bibinfo{year}{2021}{\natexlab{a}}),
  \eprint{2102.10874}.

\bibitem[{\citenamefont{Aad et~al.}(2021{\natexlab{b}})}]{Aad:2020arf}
\bibinfo{author}{\bibfnamefont{G.}~\bibnamefont{Aad}} \bibnamefont{et~al.}
  (\bibinfo{collaboration}{ATLAS}), \bibinfo{journal}{JHEP}
  \textbf{\bibinfo{volume}{02}}, \bibinfo{pages}{226}
  (\bibinfo{year}{2021}{\natexlab{b}}), \eprint{2011.05259}.

\end{thebibliography}

\end{document}